\documentclass[12pt,cite,a4paper]{article}

\usepackage{amsmath}
\usepackage{amssymb}
\usepackage[dvips]{graphicx}
\numberwithin{equation}{section}
\renewcommand{\thefootnote}{\fnsymbol{footnote}}

\begin{document}

\begin{titlepage}
\begin{flushright}
IFUP--TH/2006\,--\,7 \\
\end{flushright}
~

\vskip .8truecm
\begin{center}
\Large\bf
Liouville field theory with heavy charges.\\
II. The conformal boundary case\footnote[1]{This work is supported
in part by M.I.U.R.}
\end{center}

\vskip 2truecm
\begin{center}
{Pietro Menotti} \\
{\small\it Dipartimento di Fisica dell'Universit{\`a}, Pisa 56100,
Italy and}\\
{\small\it INFN, Sezione di Pisa}\\
{\small\it e-mail: menotti@df.unipi.it}\\
\end{center}
\vskip .8truecm
\begin{center}
{Erik Tonni} \\
{\small\it Scuola Normale Superiore, Pisa 56100, Italy and}\\
{\small\it INFN, Sezione di Pisa}\\
{\small\it e-mail: e.tonni@sns.it}\\
\end{center}

\vskip 2truecm

\begin{abstract}

We develop a general technique for computing functional integrals
with fixed area and boundary length constraints. The correct
quantum dimensions for the vertex functions are recovered by
properly regularizing the Green function. Explicit computation is
given for the one point function providing the first one loop
check of the bootstrap formula.

\end{abstract}

\vskip 1truecm

\end{titlepage}
\renewcommand{\thefootnote}{\arabic{footnote}}

\section*{Introduction}

In a preceding paper \cite{MTpseudosphere}, denoted in the
following by (I), we developed the heavy charge approach to the
correlation functions in Liouville theory on the pseudosphere.
Here we extend the treatment to the richer case of Liouville
theory on a finite domain with conformally invariant boundary
conditions. The bootstrap approach to such a problem was developed
in the seminal papers by Fateev, Zamolodchikov and Zamolodchikov
\cite{FZZ} and Teschner \cite{Teschner: remarks} providing several
profound results; in particular the exact bulk one point function
and the boundary two point function were derived. Further results
were obtained in \cite{PonsotTeschner, Hosomichi}. As done in (I)
for the pseudosphere, here we want to approach the problem in the
standard way of quantum field theory, i.e. by computing first a
stable classical background and then integrating
over the quantum fluctuations.\\
In section \ref{Boundary LFT section} we separate the action into
the classical and the quantum part
and we derive the boundary conditions for the Green function.\\
In section \ref{constrained pathint section} we develop the
technique for computing the constrained path integrals by
explicitly extracting the contribution of the fixed area and fixed
boundary length constraints. Then we consider the transformation
properties of the constrained $N$ point vertex correlation
functions under general conformal transformations. The key role of
such development is played by the regularized value of the Green
function at coincident points, both in the bulk and on the
boundary. The non invariant regularization of the Green function
suggested by Zamolodchikov and Zamolodchikov in the case of the
pseudosphere \cite{ZZpseudosphere,MTgeometric,MTtetrahedron} and
its generalization to the boundary are essential. We prove that
the one loop contribution (the quantum determinant) provides
the correct quantum dimensions \cite{CT} to the vertex operators.\\
In section \ref{one point section} we deal with the computation of
the one point function. The background generated by a single
charge is stable only in presence of a negative boundary
cosmological constant; we compute the Green function on such a
background satisfying the correct conformally invariant boundary
conditions by explicitly resumming a Fourier series, as a more
straightforward alternative to the general method employed in (I)
for the pseudosphere. Such a Green function and its regularized
value at coincident points are given in terms
of the incomplete Beta function.\\
The presence of a negative boundary cosmological constant imposes
to work with some constraints and the fixed boundary length
constraint is the most natural one. It is proved that the fixed
boundary length constraint is sufficient to make the functional
integral well defined because the operator whose determinant
provides the one loop contribution to the semiclassical result
possesses one and only one negative eigenvalue. However, to
compare our results with the ones given in \cite{FZZ} at fixed
area $A$ and fixed boundary length $l$, we introduce also the
fixed area constraint. Exploiting the decomposition found in
section \ref{constrained pathint section}, we are left with the
computation of an unconstrained functional determinant, which we
determine through the technique of varying the charges and the
invariant ratio $A/l^2$.\\
The one loop result obtained in this way agrees with the expansion
of the fixed area and boundary length one point function derived
through the bootstrap method in \cite{FZZ} and
for which there was up to now no perturbative check.\\
In appendix \ref{spectrum} we analyze the spectrum of the operator
occurring in the quantum determinant.

\section{Boundary Liouville field theory}
\label{Boundary LFT section}

\noindent The action on a finite simply connected domain $\Gamma$
with background metric $g_{ab}=\delta_{ab}$ in absence of sources
\cite{FZZ,PonsotTeschner} is
\begin{equation}
\hspace{-1cm} S_{\,\Gamma,\,0}[ \,\phi\,] \,=\,
 \int_{\Gamma_{\varepsilon}}
\left[ \,\frac{1}{\pi} \,\partial_\zeta \phi
\,\partial_{\bar{\zeta}}\phi+\mu\, e^{2b\phi}\,\right] d ^2 \zeta
\,+\, \oint_{\partial\Gamma} \left[ \,\frac{Q\,k}{2\pi}\,\phi
+\mu_{\scriptscriptstyle\hspace{-.05cm}B}\, e^{b\phi}\,\right]
d\lambda\,
\end{equation}
and in presence of sources it goes over to
\begin{eqnarray}\label{action bound Gamma}
\hspace{-1cm} S_{\,\Gamma,\,N}[ \,\phi\,] & = &
 \lim_{\varepsilon_n\,\rightarrow\, 0}\,
 \Bigg\{ \int_{\Gamma_{\varepsilon}}
\left[ \,\frac{1}{\pi} \,\partial_\zeta \phi
\,\partial_{\bar{\zeta}}\phi+\mu\, e^{2b\phi}\,\right] d ^2 \zeta
\,+\, \oint_{\partial\Gamma} \left[ \,\frac{Q\,k}{2\pi}\,\phi
+\mu_{\scriptscriptstyle\hspace{-.05cm}B}\,
e^{b\phi}\,\right] d\lambda\,  \\
  &  & \hspace{1.4cm}\rule{0pt}{.9cm}
   -\,\frac{1}{2\pi i}\;\sum_{n=1} ^N
   \alpha_n\oint_{\partial\gamma_n} \phi
\left( \, \frac{d \zeta}{\rule{0pt}{.4cm}\zeta-\zeta_n}- \frac{d
\bar{\zeta}}{\rule{0pt}{.4cm}\bar{\zeta}-\bar{\zeta}_n}\, \right)-
\sum_{n=1} ^N \alpha_n^2 \log \varepsilon_n^2 \; \Bigg\} \nonumber
\end{eqnarray}
where $Q=1/b + b$, $k$ is the extrinsic curvature of the boundary
$\partial\Gamma$, defined as
\begin{equation}\label{k definition}
k\,=\,\frac{1}{2i}\;\frac{d}{d\lambda}
\left(\log\frac{d\zeta}{d\lambda}\,-\,
\log\frac{d\bar{\zeta}}{d\lambda}\,\right)
 \hspace{1,1cm},\hspace{1,1cm} \zeta(\lambda)\in\partial\Gamma
\end{equation}
where $\lambda$ is the parametric boundary length, i.e.
$d\lambda=\sqrt{\rule{0pt}{.386cm}d\zeta d\bar{\zeta}}$. The
integration domain $\Gamma_{\varepsilon}=\Gamma\hspace{-.07cm}
\setminus \bigcup_{n=1}^N \gamma_{n}$ is obtained by removing $N$
infinitesimal disks $\gamma_{n}=\{|\zeta-\zeta_n|<\varepsilon_n\}$
from the simply connected domain $\Gamma$.\\
The boundary behavior of $\phi$ near the sources is
\begin{equation}
\label{phi bc sources} \phi(\zeta)  \,=\, -\,\alpha_n  \log
|\zeta-\zeta_n|^2\,+O(1)
\hspace{.9cm}\textrm{when}\hspace{.9cm}\,\zeta\,\rightarrow
\zeta_n\;.
\end{equation}

\noindent In order to connect the quantum theory to its
semiclassical limit it is useful to define \cite{ZZsphere}
\begin{equation}
\varphi \,=\, 2 b \phi \hspace{.8cm},\hspace{.8cm}
\alpha_n\,=\,\frac{\eta_n}{b}\;.
\end{equation}
Then, we decompose the field $\varphi$ as the sum of a classical
background field $\varphi_{\scriptscriptstyle\hspace{-.05cm}B}$
and a quantum field
\begin{equation}\label{phi decomposition BLFT}
\varphi \,=\,\varphi_{\scriptscriptstyle\hspace{-.05cm}B} +
2b\,\chi\;.
\end{equation}
The condition of local finiteness of the area around each source
and the asymptotic behavior
(\ref{phi bc sources}) for the field $\phi$ imposes that
$1-2 \eta_n>0$ \cite{Seiberg: Notes, Picard,GinspargMoore}.\\
Then, we can write the action as the sum of a classical and a
quantum action as follows
\begin{equation}\label{action BLFT splitting}
S_{\,\Gamma,\,N}[ \,\phi\,]  \,=\,S_{cl}[
\,\varphi_{\scriptscriptstyle\hspace{-.05cm}B}\,] +
S_{q}[\,\varphi_{\scriptscriptstyle\hspace{-.05cm}B},\,\chi\,]\;.
\end{equation}
The classical action in absence of sources is given by
\begin{equation}
S_{cl,\,0}[ \,\varphi_{\scriptscriptstyle\hspace{-.05cm}B}\,] \, =
\,\frac{1}{b^2}\;\Bigg\{ \int_{\Gamma} \left[ \,\frac{1}{4\pi}
\,\partial_\zeta \varphi_{\scriptscriptstyle\hspace{-.05cm}B}
\,\partial_{\bar{\zeta}}\varphi_{\scriptscriptstyle\hspace{-.05cm}B}
+\mu
b^2\, e^{\varphi_{\scriptscriptstyle\hspace{-.02cm}B}}\,\right] d
^2 \zeta \,+\, \oint_{\partial\Gamma} \left[
\,\frac{k}{4\pi}\,\varphi_{\scriptscriptstyle\hspace{-.05cm}B}
+\mu_{\scriptscriptstyle\hspace{-.05cm}B}b^2\,
e^{\varphi_{\scriptscriptstyle\hspace{-.02cm} B}/2}\,\right]
d\lambda\,\Bigg\}
\end{equation}
and in presence of sources it goes over to
\begin{eqnarray}\label{action classical phiB BLFT}
\hspace{0cm} S_{cl}[
\,\varphi_{\scriptscriptstyle\hspace{-.05cm}B}\,] & =
&\frac{1}{b^2}\;
 \lim_{\varepsilon_n\,\rightarrow\, 0}\,
 \Bigg\{ \int_{\Gamma_{\varepsilon}}
\left[ \,\frac{1}{4\pi} \,\partial_\zeta
\varphi_{\scriptscriptstyle\hspace{-.05cm}B}
\,\partial_{\bar{\zeta}}\varphi_{\scriptscriptstyle\hspace{-.05cm}B}
+\mu
b^2\, e^{\varphi_{\scriptscriptstyle\hspace{-.02cm}B}}\,\right] d
^2 \zeta + \oint_{\partial\Gamma} \left[
\,\frac{k}{4\pi}\,\varphi_{\scriptscriptstyle\hspace{-.05cm}B}
+\mu_{\scriptscriptstyle\hspace{-.05cm}B}b^2\,
e^{\varphi_{\scriptscriptstyle\hspace{-.02cm} B}/2}\,\right]
d\lambda\,  \nonumber\\
  &  & \hspace{2cm}\rule{0pt}{.9cm}
   -\,\frac{1}{4\pi i}\;\sum_{n=1} ^N
   \eta_n\oint_{\partial\gamma_n}\hspace{-.2cm}
   \varphi_{\scriptscriptstyle\hspace{-.05cm}B}
\left( \, \frac{d \zeta}{\rule{0pt}{.4cm}\zeta-\zeta_n}- \frac{d
\bar{\zeta}}{\rule{0pt}{.4cm}\bar{\zeta}-\bar{\zeta}_n}\, \right)-
\sum_{n=1} ^N \eta_n^2 \log \varepsilon_n^2 \; \Bigg\}
\end{eqnarray}
while the quantum action reads
\begin{eqnarray}\label{quantum action BLFT first}
\hspace{0cm} S_{q}[
\,\varphi_{\scriptscriptstyle\hspace{-.05cm}B}\,, \chi\,] & = &
 \lim_{\varepsilon_n\,\rightarrow\, 0}\,
\Bigg\{ \int_{\Gamma_{\varepsilon}} \left[ \,\frac{1}{\pi}
\,\partial_\zeta \chi \,\partial_{\bar{\zeta}}\chi+\mu\,
e^{\,\varphi_{\hspace{-.02cm}
\scriptscriptstyle{B}}}\,\big(\,e^{2b\,\chi}-1\,\big)
-\frac{1}{\pi b}\;\chi\,\partial_\zeta\partial_{\bar{\zeta}}
\varphi_{\scriptscriptstyle\hspace{-.05cm}B}\,\right]\, d^2 \zeta
 \\
\rule{0pt}{.8cm} & & \hspace{-.4cm}+\,\frac{1}{4\pi i\,b}\;
\oint_{\partial\Gamma}
  \chi\;
\Big( \,
\partial_\zeta \varphi_{\scriptscriptstyle\hspace{-.05cm}B}\,d\zeta-
\partial_{\bar{\zeta}}
\varphi_{\scriptscriptstyle\hspace{-.05cm}B}\,d\bar{\zeta} \,\Big)
\,+\, \oint_{\partial\Gamma} \left[ \,\frac{Q\,k}{2\pi}\,\chi
+\mu_{\scriptscriptstyle\hspace{-.05cm}B}\,
e^{\varphi_{\scriptscriptstyle\hspace{-.02cm}B}/2}\,
\big(\,e^{b\,\chi}-1\,\big)\,\right]
d\lambda\,
 \nonumber \\
\rule{0pt}{.9cm} & & \hspace{1cm}
+\,\frac{1}{4\pi}\oint_{\partial\Gamma}k\;
\varphi_{\scriptscriptstyle\hspace{-.05cm}B}\,d\lambda\,
-\frac{1}{2\pi i\,b}\; \sum_{n=1} ^N \;\oint_{\partial\gamma_n}
\chi\, \left( \, \frac{\eta_n}{\rule{0pt}{.4cm}\zeta-\zeta_n}
\,+\, \frac{1}{2}\,\partial_\zeta
\varphi_{\scriptscriptstyle\hspace{-.05cm}B}\right)\,d\zeta
 \nonumber \\
 \rule{0pt}{.9cm}   &  &
\hspace{4.3cm} +\,\frac{1}{2\pi i\,b}\; \sum_{n=1} ^N\;
\oint_{\partial\gamma_n}   \chi\, \left( \,
\frac{\eta_n}{\rule{0pt}{.4cm}\bar{\zeta}-\bar{\zeta}_n}\,+\,
\frac{1}{2}\,\partial_{\bar{\zeta}}
\varphi_{\scriptscriptstyle\hspace{-.05cm}B}\right)\,d\bar{\zeta}
       \;\Bigg\}\;.\nonumber
\end{eqnarray}
For the classical background field, we assume the following
boundary behavior
\begin{equation}
\label{varphiB bc sources BLFT}
\varphi_{\scriptscriptstyle\hspace{-.05cm}B}(\zeta)  \,=\,
-\;2\eta_n \log \,|\,\zeta-\zeta_n\,|^2\,+O(1)
\qquad\hspace{.7cm}\textrm{when}\hspace{.7cm}\qquad\,\zeta\,
\rightarrow
\zeta_n\;.
\end{equation}
Under a generic conformal transformation $\zeta \rightarrow
\tilde{\zeta}=\tilde{\zeta}(\zeta)$ the background field changes
as follows
\begin{equation}\label{varphiB transformations BLFT}
\varphi_{\scriptscriptstyle\hspace{-.05cm}B}(\zeta)
\hspace{.5cm}\longrightarrow  \hspace{.5cm}
\tilde{\varphi}_{\scriptscriptstyle\hspace{-.05cm}B}(\tilde{\zeta})
\;=\;\varphi_{\scriptscriptstyle\hspace{-.05cm}B}(\zeta)\,-\, \log
\bigg| \frac{d \tilde{\zeta}}{d \zeta}\bigg|^{2}
\end{equation}
so that $e^{\varphi_{\scriptscriptstyle\hspace{-.05cm}B}}d^2\zeta$
is invariant, while the extrinsic curvature becomes
\begin{equation}\label{curvature transformation}
k\hspace{.5cm}\longrightarrow\hspace{.5cm}\tilde{k}\,=\,
\frac{1}{\sqrt{J}}\left(k\,+\,\frac{1}{2i}
\left(\frac{d\zeta}{d\lambda}\,\partial_\zeta\log
J\,-\,\frac{d\bar{\zeta}}{d\lambda}\,\partial_{\bar{\zeta}}\log
J\right)\right) \hspace{.6cm},
\hspace{.6cm}\zeta(\lambda)\in\partial\Gamma
\end{equation}
where $J\equiv|d\tilde{\zeta}/d\zeta|^2$. Under such conformal
transformations the classical action both in absence and in
presence of sources is invariant up to a field independent term.
Thus, the classical action (\ref{action classical phiB BLFT}) by
variation of the field
$\varphi_{\scriptscriptstyle\hspace{-.05cm}B}$ gives rise to the
conformally invariant field equation
\begin{equation}\label{liouville eq with sources varphi BLFT}
    -\,\partial_{\zeta}\partial_{\bar{\zeta}}\,
    \varphi_{\scriptscriptstyle\hspace{-.05cm}B}\,
    +
    2\pi \,\mu b^2 \,e^{\,\varphi_{\hspace{-.02cm}
    \scriptscriptstyle{B}}} \,=\,
    2\pi \sum_{n=1}^N \eta_n
    \,\delta^2(\zeta-\zeta_n)\;
\end{equation}
which is the Liouville equation in presence of $N$ sources, and to
the following conformally invariant boundary conditions for the
classical field
\begin{equation}\label{FZZ b.c. classic field}
-\,\frac{1}{4\pi i}\,\left( \,
\frac{d\zeta}{d\lambda}\,\partial_\zeta
\varphi_{\scriptscriptstyle\hspace{-.05cm}B}-
\frac{d\bar{\zeta}}{d\lambda}\,\partial_{\bar{\zeta}}
\varphi_{\scriptscriptstyle\hspace{-.05cm}B} \right)\,=\,
\frac{k}{2\pi}\,+\,\mu_{\scriptscriptstyle\hspace{-.05cm}B}b^2\,
e^{\varphi_{\scriptscriptstyle\hspace{-.02cm}B}/2}
\hspace{.7cm},\hspace{.7cm} \zeta(\lambda)\in\partial\Gamma\;.
\end{equation}
The field independent terms which appear in the change of the
actions under a conformal transformation are
\begin{equation}\label{action BLFT transf law classic no sources}
\widetilde{S}_{cl,\,0}[
\,\tilde{\varphi}_{\scriptscriptstyle\hspace{-.05cm}B}\,]\,=\,
S_{cl,\,0}[
\,\varphi_{\scriptscriptstyle\hspace{-.05cm}B}\,]\,+\,\frac{1}{8\pi
b^2}\left(\,
\oint_{\partial\widetilde{\Gamma}}\tilde{k}\,\log\tilde{J}
\,d\tilde{\lambda}\,-\,\oint_{\partial\Gamma}k\,\log J
\,d\lambda\right)
\end{equation}
where $\tilde{J}=|d\zeta/d\tilde{\zeta}|^2=1/J$, while in presence
of sources we have
\begin{eqnarray}\label{action BLFT transf law classic}
\widetilde{S}_{cl}[
\,\tilde{\varphi}_{\scriptscriptstyle\hspace{-.05cm}B}\,]&=&
S_{cl}[ \,\varphi_{\scriptscriptstyle\hspace{-.05cm}B}\,] \,+\,
\sum_{n\,=\,1}^N  \frac{\eta_n (\,1-\eta_n)}{b^2}\; \log\bigg|
\frac{d \tilde{\zeta}}{d \zeta}
\bigg|_{\zeta\,=\,\zeta_n}^2\\
\rule{0pt}{.7cm}& & \hspace{2cm}+\,\frac{1}{8\pi b^2}\left(\,
\oint_{\partial\widetilde{\Gamma}}\tilde{k}\,\log\tilde{J}
\,d\tilde{\lambda}\,-\,\oint_{\partial\Gamma}k\,\log J
\,d\lambda\right).\nonumber
\end{eqnarray}
The requirement that the expectation value of 1 be invariant under
conformal transformations, i.e. the invariance of the vacuum,
imposes to subtract the term
\begin{equation}\label{extra terms}
\frac{1}{8\pi b^2}\left(\,
\oint_{\partial\widetilde{\Gamma}}\tilde{k}\,\log\tilde{J}
\,d\tilde{\lambda}\,-\,\oint_{\partial\Gamma}k\,\log J
\,d\lambda\right)
\end{equation}
from the r.h.s. of (\ref{action BLFT transf law classic no
sources}) and (\ref{action BLFT transf law classic}) when
computing the transformation of the vertex correlation functions
under conformal transformations. The term (\ref{extra terms})
vanishes identically for the conformal transformations which map
the unit disk into itself, i.e. the $SU(1,1)$ transformations. In
this way one obtains the semiclassical conformal dimensions of the
vertex operators $e^{2(\eta_n/b)\phi(\zeta_n)}$
\begin{equation}
\frac{\eta_n(1-\eta_n)}{b^2}\,=\,\alpha_n
\left(\,\frac{1}{b}-\alpha_n\right)\;.
\end{equation}
We recall that $\mu b^2$ and
$\mu_{\scriptscriptstyle\hspace{-.05cm}B}b^2$
have to be kept constant when $b \rightarrow 0$
\cite{FZZ, TeschnerReview}.\\
Using the equation of motion for the classical field, the boundary
conditions (\ref{FZZ b.c. classic field}) and the behavior at the
sources (\ref{varphiB bc sources BLFT}), the quantum action
(\ref{quantum action BLFT first}) becomes
\begin{eqnarray}\label{quantumaction BLFT}
\hspace{0cm} S_{q}[
\,\varphi_{\scriptscriptstyle\hspace{-.05cm}B}\,, \chi\,] &=&
\int_{\Gamma} \left[ \;\frac{1}{\pi} \,\partial_\zeta \chi
\,\partial_{\bar{\zeta}}\chi+\mu\, e^{\,\varphi_{\hspace{-.02cm}
\scriptscriptstyle{B}}}\,\big(\,e^{2b\,\chi}-1-2b\chi\,\big)\,\right]\,
d^2 \zeta
 \\
\rule{0pt}{.9cm} & & \hspace{.4cm}
+\,\frac{1}{4\pi}\,\oint_{\partial\Gamma}k\;
\varphi_{\scriptscriptstyle\hspace{-.05cm}B}\,d\lambda\,
+\,\frac{b}{2\pi}\, \oint_{\partial\Gamma}k\;\chi\,d\lambda \,+\,
\oint_{\partial\Gamma} \mu_{\scriptscriptstyle\hspace{-.05cm}B}\,
e^{\varphi_{\scriptscriptstyle\hspace{-.02cm}B}/2}\,
\big(\,e^{b\,\chi}-1-b\chi\,\big)\, d\lambda\;.\nonumber
\end{eqnarray}
Integrating by parts the volume integral in (\ref{quantumaction
BLFT}) we obtain
\begin{eqnarray}\label{quantumaction BLFT 2}
\hspace{0cm} S_{q}[
\,\varphi_{\scriptscriptstyle\hspace{-.05cm}B}\,, \chi\,] &=&
\int_{\Gamma} \left[ \;-\,\frac{1}{\pi}\;\chi
\,\partial_\zeta\partial_{\bar{\zeta}}\,\chi+\mu\,
e^{\,\varphi_{\hspace{-.02cm}
\scriptscriptstyle{B}}}\,\big(\,e^{2b\,\chi}-1-2b\chi\,\big)
\,\right]\,
d^2 \zeta \\
\rule{0pt}{.9cm} & & \hspace{.4cm} +\,\frac{1}{4\pi i}\,
\oint_{\partial\Gamma}\chi\left( \,
\frac{d\zeta}{d\lambda}\,\partial_\zeta \chi-
\frac{d\bar{\zeta}}{d\lambda}\,\partial_{\bar{\zeta}} \chi \right)
\nonumber\\
\rule{0pt}{.9cm} & & \hspace{.4cm}
+\,\frac{1}{4\pi}\,\oint_{\partial\Gamma}k\;
\varphi_{\scriptscriptstyle\hspace{-.05cm}B}\,d\lambda\,
+\,\frac{b}{2\pi}\, \oint_{\partial\Gamma}k\;\chi\,d\lambda \,+\,
\oint_{\partial\Gamma} \mu_{\scriptscriptstyle\hspace{-.05cm}B}\,
e^{\varphi_{\scriptscriptstyle\hspace{-.02cm}B}/2}\,
\big(\,e^{b\,\chi}-1-b\chi\,\big)\, d\lambda\;.\nonumber
\end{eqnarray}
By expanding in $b$ the boundary conditions for the full field
$\varphi=\varphi_{\scriptscriptstyle\hspace{-.05cm}B}+2b\chi\,$,
which are
\begin{equation}\label{FZZ b.c. varphi}
-\,\frac{1}{4\pi i}\,\left( \,
\frac{d\zeta}{d\lambda}\,\partial_\zeta \varphi\,-\,
\frac{d\bar{\zeta}}{d\lambda}\,\partial_{\bar{\zeta}} \varphi
\right)\,=\,
\frac{Q\,k}{2\pi}\,b\,+\,\mu_{\scriptscriptstyle\hspace{-.05cm}B}b^2\,
e^{\varphi/2} \hspace{.7cm},\hspace{.7cm}
\zeta(\lambda)\in\partial\Gamma
\end{equation}
and using the boundary conditions (\ref{FZZ b.c. classic field})
for the classical background field
$\varphi_{\scriptscriptstyle\hspace{-.05cm}B}$ extracted from the
classical action (\ref{action classical phiB BLFT}), we get the
boundary conditions for $\chi$
\begin{eqnarray}\label{FZZ full b.c. chi}
\hspace{-1.2cm} -\,\frac{1}{2\pi i}\,\left( \,
\frac{d\zeta}{d\lambda}\,\partial_\zeta \chi\,-\,
\frac{d\bar{\zeta}}{d\lambda}\,\partial_{\bar{\zeta}} \chi \right)
&=& \mu_{\scriptscriptstyle\hspace{-.05cm}B}b\,
e^{\varphi_{\scriptscriptstyle\hspace{-.02cm}B}/2}\,
\big(\,e^{b\chi}-1\,\big)\,+\,\frac{k}{2\pi}\,b
\hspace{.7cm},\hspace{.7cm}
\zeta(\lambda)\in\partial\Gamma\\
\rule{0pt}{.8cm} &=& \mu_{\scriptscriptstyle\hspace{-.05cm}B}b^2\,
e^{\varphi_{\scriptscriptstyle\hspace{-.02cm}B}/2}\,\chi + b
\left(\,\frac{k}{2\pi}+\mu_{\scriptscriptstyle\hspace{-.05cm}B}b^2\,
e^{\varphi_{\scriptscriptstyle\hspace{-.02cm}B}/2}\,\frac{\chi^2}{2}\,
\right)+O(b^2)\;.\nonumber
\end{eqnarray}
To order $O(b^0)$ we have
\begin{equation}\label{FZZ b.c. chi}
-\,\frac{1}{2\pi i}\,\left( \,
\frac{d\zeta}{d\lambda}\,\partial_\zeta \chi\,-\,
\frac{d\bar{\zeta}}{d\lambda}\,\partial_{\bar{\zeta}} \chi
\right)\,=\, \mu_{\scriptscriptstyle\hspace{-.05cm}B}b^2\,
e^{\varphi_{\scriptscriptstyle\hspace{-.02cm}B}/2}\,
\chi\hspace{.7cm},\hspace{.7cm}
\zeta(\lambda)\in\partial\Gamma\;.
\end{equation}
With the field $\chi$ satisfying (\ref{FZZ b.c. chi}), we are left
with the following quantum action
\begin{eqnarray}\label{quantumaction BLFT 3}
\hspace{-1.2cm} S_{q}[
\,\varphi_{\scriptscriptstyle\hspace{-.05cm}B}\,, \chi\,] &=&
\frac{1}{2}\,\int_{\Gamma} \chi \left( -\frac{2}{\pi}
\,\partial_\zeta\partial_{\bar{\zeta}}\,+\,4\,\mu b^2\,
e^{\,\varphi_{\hspace{-.02cm}
\scriptscriptstyle{B}}}\,\right)\chi\, d^2
\zeta\,+\,\sum_{k\,\geqslant\,3}\int_{\Gamma}\mu\,
e^{\,\varphi_{\hspace{-.02cm} \scriptscriptstyle{B}}}
\;\frac{(2b\chi)^k}{k!}\,d^2z \nonumber\\
\rule{0pt}{.9cm} & & +\,\frac{1}{4\pi}\,\oint_{\partial\Gamma}k\;
\varphi_{\scriptscriptstyle\hspace{-.05cm}B}\,d\lambda\,
+\,\frac{b}{2\pi}\, \oint_{\partial\Gamma}k\;\chi\,d\lambda\,+\,
\sum_{k\,\geqslant\,3}\oint_{\partial\Gamma}
\mu_{\scriptscriptstyle\hspace{-.05cm}B}\,
e^{\,\varphi_{\hspace{-.02cm} \scriptscriptstyle{B}}/2}
\;\frac{(b\chi)^k}{k!}\,d\lambda\;.
\end{eqnarray}
The first term of the second line is $O(b^0)$ while the other
boundary terms are $O(b)$ or higher order in $b$. \\
Thus, imposing on the Green function $g(\zeta,\zeta')$ of the
following operator
\begin{equation}
D\,\equiv\,-\,\frac{2}{\pi}
\,\partial_\zeta\partial_{\bar{\zeta}}\,+\,4\,\mu b^2\,
e^{\,\varphi_{\hspace{-.02cm} \scriptscriptstyle{B}}}
\end{equation}
the mixed boundary conditions (\ref{FZZ b.c. chi}), i.e.
\begin{equation}\label{FZZ b.c. propagator Gamma}
-\,\frac{1}{2\pi i}\,\left( \,
\frac{d\zeta}{d\lambda}\,\partial_\zeta \,g(\zeta,\zeta')\,-\,
\frac{d\bar{\zeta}}{d\lambda}\,\partial_{\bar{\zeta}}\,
g(\zeta,\zeta') \right)\,=\,
\mu_{\scriptscriptstyle\hspace{-.05cm}B}b^2\,
e^{\varphi_{\scriptscriptstyle\hspace{-.02cm}B}/2}\,
g(\zeta,\zeta')\hspace{.6cm},\hspace{.6cm}
\zeta(\lambda)\in\partial\Gamma
\end{equation}
we can develop a perturbative expansion in $b$. The Green function
of the operator $D$ satisfies
\begin{equation}
D\,g(\zeta,\zeta')\,=\,\delta^2(\zeta-\zeta')
\end{equation}
and, due to the covariance of $D$ and of the boundary conditions
(\ref{FZZ b.c. propagator Gamma}), it is invariant in value under
a conformal transformation
$\zeta\rightarrow\tilde{\zeta}=\tilde{\zeta}(\zeta)$, i.e.
\begin{equation}
\tilde{g}(\tilde{\zeta},\tilde{\zeta}')\,=\,g(\zeta,\zeta')\;.
\end{equation}

\section{Constrained path integral and quantum dimensions}
\label{constrained pathint section}

The partition function in presence of sources is given by
\begin{equation}\label{N point geometric}
Z(\zeta_1,\eta_1,\dots,\zeta_N,\eta_N;\mu,
\mu_{\scriptscriptstyle\hspace{-.05cm}B})\,=\,
\int\hspace{-.06cm} \mathcal{D}\,[\, \phi \,]\;\,
e^{\,-S_{\Gamma,N}[\,\phi\,]}
\end{equation}
with
\begin{equation}
Z(\zeta_1,\eta_1,\dots,\zeta_N,\eta_N;\mu,
\mu_{\scriptscriptstyle\hspace{-.05cm}B})\,\equiv\,
\int_0^\infty\frac{dl}{l}\;
e^{-\mu_{\scriptscriptstyle\hspace{-.05cm}B}l}
\int_0^\infty\frac{dA}{A}\;e^{-\mu A}\,
Z(\zeta_1,\eta_1,\dots,\zeta_N,\eta_N;A,l\hspace{.04cm})
\end{equation}
where we have used the conventions of \cite{FZZ} and
\begin{eqnarray}\label{fixed A and l partfunc}
Z(\zeta_1,\eta_1,\dots,\zeta_N,\eta_N;A,l\hspace{.04cm})&=&
e^{-S_{cl}[\varphi_{\scriptscriptstyle\hspace{-.03cm}B}]}
\;A\,l\int\hspace{-.06cm} \mathcal{D}\,[\, \chi \,]\;\,
e^{\,-S_{q}[\,\chi,\varphi_{\scriptscriptstyle\hspace{-.03cm}B}\,]}\;
\times\\
\rule{0pt}{.7cm}& & \hspace{1.4cm}\times\;
\delta\left(\int_{\Gamma}
e^{\varphi_{\scriptscriptstyle\hspace{-.03cm}B}+2b\chi}d^2\zeta-A\right)
\delta\left(\oint_{\partial\Gamma}
e^{\varphi_{\scriptscriptstyle\hspace{-.03cm}B}/2+b\chi}
    d\lambda-l\right)\,.\nonumber
\end{eqnarray}
The classical background field
$\varphi_{\scriptscriptstyle\hspace{-.05cm}B}$ satisfies the
Liouville equation (\ref{liouville eq with sources varphi BLFT})
with boundary conditions (\ref{FZZ b.c. classic field}) and
\begin{eqnarray}\label{area varphiB}
   A \hspace{-.1cm} &=& \hspace{-.2cm}\int_{\Gamma}
   e^{\varphi_{\scriptscriptstyle\hspace{-.03cm}B}}
    d^2\zeta\\
    \label{length varphiB}
\rule{0pt}{.8cm}l \hspace{-.1cm}& = & \hspace{-.2cm}
\oint_{\partial\Gamma}
e^{\varphi_{\scriptscriptstyle\hspace{-.03cm}B}/\,2}\,
    d\lambda\;.
\end{eqnarray}
Substituting (\ref{quantumaction BLFT 3}) in (\ref{fixed A and l
partfunc}) and exploiting (\ref{area varphiB}) and (\ref{length
varphiB}), we have to one loop
\begin{equation}\label{fixed A and l one loop}
Z(\zeta_1,\eta_1,\dots,\zeta_N,\eta_N;A,l\,)\,=\,
e^{-S_{cl}[\varphi_{\scriptscriptstyle\hspace{-.03cm}B}]}\;
\frac{A\,l}{2b^2} \;I
\end{equation}
where
\begin{equation}\label{fixed A and l one loop bis}
I\,\equiv\,e^{-\frac{1}{4\pi}\oint_{\partial\Gamma}k\,
\varphi_{\scriptscriptstyle\hspace{-.05cm}B}\,d\lambda}
\int\hspace{-.06cm} \mathcal{D}\,[\, \chi \,]\;\,
e^{\,-\frac{1}{2}(\chi,D\chi)}\;
\delta\left(\int_{\Gamma}
e^{\varphi_{\scriptscriptstyle\hspace{-.03cm}B}}\chi\,d^2\zeta\right)
\delta\left(\oint_{\partial\Gamma}
e^{\varphi_{\scriptscriptstyle\hspace{-.03cm}B}/2}\chi\,d\lambda\right)\;.
\end{equation}
The seemingly non perturbative factor $1/b^2$ in (\ref{fixed A and
l one loop}) is due to the presence of the constraints.\\
Using the integral representation for the two delta functions
\cite{Ambjorn} we have
\begin{eqnarray}
I&=& e^{-\frac{1}{4\pi}\oint_{\partial\Gamma}k\,
\varphi_{\scriptscriptstyle\hspace{-.05cm}B}\,d\lambda}\,\times\\
\rule{0pt}{.8cm}& &\hspace{-.66cm}
\times\,\frac{1}{(2\pi)^2}\int\hspace{-.06cm}\mathcal{D}\,[\,\chi\,]\,
\int d\rho\int d\tau\;
\exp\left\{-\frac{1}{2}\,\big(\chi,D\chi\big) +
i\,\rho\int_{\Gamma}e^{\varphi_{\scriptscriptstyle\hspace{-.03cm}B}}
\chi\,d^2\zeta
+i\,\tau\oint_{\partial\Gamma}
e^{\varphi_{\scriptscriptstyle\hspace{-.03cm}B}/2}
\chi\,d\lambda\,\right\}.
\nonumber
\end{eqnarray}
In the following we shall use the notation
$\varphi_{\scriptscriptstyle\hspace{-.03cm}B}(\lambda)$ to denote
the field $\varphi_{\scriptscriptstyle\hspace{-.03cm}B}$ computed
at the boundary point $\zeta(\lambda)\in\partial\Gamma$ and
$g(\zeta,\lambda)$ and $g(\lambda,\lambda')$ to denote the values
of the Green function with one or two arguments on the
boundary.\\
Performing the field translation
\begin{equation}
\chi(\zeta)\,=\,\chi'(\zeta)+
i\,\rho\int_{\Gamma}g(\zeta,\zeta')\,
e^{\varphi_{\scriptscriptstyle\hspace{-.03cm}B}(\zeta')}d^2\zeta'
+i\,\tau\oint_{\partial\Gamma}g(\zeta,\lambda)\,
e^{\varphi_{\scriptscriptstyle\hspace{-.03cm}B}(\lambda)/2}d\lambda
\end{equation}
we reach the result
\begin{equation}\label{I factorized}
I\,=\,\frac{e^{-\frac{1}{4\pi}\oint_{\partial\Gamma}k\,
\varphi_{\scriptscriptstyle\hspace{-.05cm}B}\,d\lambda}
}{2\pi\sqrt{\textrm{det}M\;\textrm{Det}D}}
\end{equation}
where $M$ is the matrix
\begin{equation}
M\,=\,\left(\begin{array}{cc}
             L & R\\
             R & S
            \end{array}\right)
\end{equation}
with
\begin{eqnarray}
L&=&\oint_{\partial\Gamma}\oint_{\partial\Gamma}
e^{\varphi_{\scriptscriptstyle\hspace{-.03cm}B}(\lambda)/2}d\lambda
\;g(\lambda,\lambda')\,
d\lambda'\,e^{\varphi_{\scriptscriptstyle\hspace{-.03cm}B}(\lambda')/2} \\
\rule{0pt}{.8cm} S&=&\int_{\Gamma}\int_{\Gamma}
e^{\varphi_{\scriptscriptstyle\hspace{-.03cm}B}(\zeta)}d^2\zeta
\;g(\zeta,\zeta')\,
d^2\zeta'\,e^{\varphi_{\scriptscriptstyle\hspace{-.03cm}B}(\zeta')}\\
\rule{0pt}{.8cm} R&=&\int_{\Gamma}\oint_{\partial\Gamma}
e^{\varphi_{\scriptscriptstyle\hspace{-.03cm}B}(\zeta)}d^2\zeta
\;g(\zeta,\lambda)\,
d\lambda\,e^{\varphi_{\scriptscriptstyle\hspace{-.03cm}B}(\lambda)/2}
\end{eqnarray}
and $(\textrm{Det}D)^{-1/2}$ is the unconstrained path integral
\begin{equation}
\big(\textrm{Det}D\,\big)^{-1/2}\,=\,\int\hspace{-.06cm}
\mathcal{D}\,[\, \chi \,]\; e^{\,-\frac{1}{2}(\chi,D\chi)}
\end{equation}
with $\chi$ satisfying the boundary conditions (\ref{FZZ b.c. chi}).\\
In section (\ref{m0 sector sec}) it will be proved that the
expression (\ref{I factorized}) holds also when the operator $D$
has a finite number of negative eigenvalues, in which case
$|\textrm{Det}D|^{-1/2}$ is defined by
\begin{equation}
\prod_k\frac{\sqrt{2\pi}}{\sqrt{-\mu_k}}\,\int\hspace{-.06cm}
\mathcal{D}\,[\, \chi_\perp \,]\;
e^{\,-\frac{1}{2}(\chi_\perp,D\chi_\perp)}
\end{equation}
with $k$ running over the negative eigenvalues $\mu_k$ and
$\chi_\perp$ spans the subspace orthogonal to the eigenfunctions
of $D$ relative to the negative eigenvalues.\\

\noindent We are interested in the transformation law of
$I=I(\zeta_1,\eta_1,\dots,\zeta_N,\eta_N;A,l\,)$ under a conformal
transformation
$\zeta\rightarrow\tilde{\zeta}=\tilde{\zeta}(\zeta)$. \\
We notice that the matrix elements of $M$ are invariant under
conformal transformations; hence we have to study the
transformation properties of
\begin{equation}\label{I1 def}
I_1\,\equiv\,e^{-\frac{1}{4\pi}\oint_{\partial\Gamma}k\,
\varphi_{\scriptscriptstyle\hspace{-.05cm}B}\,
d\lambda}\int\hspace{-.06cm}
\mathcal{D}\,[\, \chi \,]\; e^{\,-\frac{1}{2}(\chi,D\chi)}\;.
\end{equation}
To this end, we consider  the eigenvalue equation
\begin{equation}\label{eigenvalueeq FZZ}
\left(-\frac{2}{\pi}\,\partial_\zeta\partial_{\bar{\zeta}} +
4\,\mu b^2\,
e^{\varphi_{\scriptscriptstyle\hspace{-.03cm}B}}\right)\chi_n
\,=\,\mu_n \,\chi_n
\end{equation}
with boundary conditions
\begin{equation}\label{FZZ b.c. chi eigenfunctions}
-\,\frac{1}{2\pi i}\,\left( \,
\frac{d\zeta}{d\lambda}\,\partial_\zeta \chi_n\,-\,
\frac{d\bar{\zeta}}{d\lambda}\,\partial_{\bar{\zeta}} \chi_n
\right)\,=\, \mu_{\scriptscriptstyle\hspace{-.05cm}B}b^2\,
e^{\varphi_{\scriptscriptstyle\hspace{-.02cm}B}/2}\,
\chi_n\hspace{.7cm},\hspace{.7cm}
\zeta(\lambda)\in\partial\Gamma\;.
\end{equation}
Taking the variation of (\ref{eigenvalueeq FZZ}), we get
\begin{equation}\label{variation eigeneq}
\left(-\frac{2}{\pi}\,\partial_z\partial_{\bar z} + 4\,
\mu b^2\,
e^{\varphi_{\scriptscriptstyle\hspace{-.03cm}B}}\right)
\delta\chi_n+\,4\,\chi_n\,\delta(\mu
b^2
e^{\varphi_{\scriptscriptstyle\hspace{-.03cm}B}})\,=\,
\delta\mu_n\,\chi_n+\mu_n\,\delta\chi_n\;.
\end{equation}
Then we multiply (\ref{variation eigeneq}) by $\chi_n$ and we
integrate the result on the domain $\Gamma$. Exploiting the
orthonormality of the eigenfunctions $\chi_n$, the eigenvalue
equation (\ref{eigenvalueeq FZZ}) and the divergence theorem, we
get
\begin{equation}\label{variation eigenvalue FZZ}
\delta\mu_n\,=\,4\int_{\Gamma}\chi_n^2\,\delta\big(\mu
b^2e^{\varphi_{\scriptscriptstyle\hspace{-.03cm}B}}\big)\,
d^2\zeta
\,-\,\frac{1}{2\pi}\oint_{\partial\Gamma}
\big(\,\chi_n\,\partial_{\hat{n}}\delta\chi_n-\,
\delta\chi_n\,\partial_{\hat{n}}\chi_n\,\big)d\lambda
\end{equation}
where $\partial_{\hat{n}}$ denotes the outward normal derivative
on the boundary
\begin{equation}
\partial_{\hat{n}}\,=\,\frac{1}{i}\left( \,
\frac{d\zeta}{d\lambda}\,\partial_\zeta \,-\,
\frac{d\bar{\zeta}}{d\lambda}\,\partial_{\bar{\zeta}}
\right)\hspace{.7cm},\hspace{.7cm}
\zeta(\lambda)\in\partial\Gamma\;.
\end{equation}
On the other hand, the variation of the boundary conditions
(\ref{FZZ b.c. chi eigenfunctions}) gives
\begin{equation}\label{variation eigen b.c.}
-\,\frac{1}{2\pi i}\,\left( \,
\frac{d\zeta}{d\lambda}\,\partial_\zeta \delta\chi_n\,-\,
\frac{d\bar{\zeta}}{d\lambda}\,\partial_{\bar{\zeta}} \delta\chi_n
\right)\,=\,\delta\big(\mu_{\scriptscriptstyle\hspace{-.05cm}B}b^2
e^{\varphi_{\scriptscriptstyle\hspace{-.03cm}B}/2}\big)\,\chi_n+
\mu_{\scriptscriptstyle\hspace{-.05cm}B}b^2
e^{\varphi_{\scriptscriptstyle\hspace{-.03cm}B}/2}\,
\delta\chi_n\hspace{.5cm},\hspace{.5cm}
\zeta(\lambda)\in\partial\Gamma\;.
\end{equation}
Using (\ref{FZZ b.c. chi eigenfunctions}) and (\ref{variation
eigen b.c.}), we find that (\ref{variation eigenvalue FZZ})
becomes
\begin{equation}\label{variation eigenvalue FZZ final}
\delta\mu_n\,=\,4\int_{\Gamma}\chi_n^2(\zeta)\,\delta\big(\mu
b^2e^{\varphi_{\scriptscriptstyle\hspace{-.03cm}B}}\big)\,d^2\zeta\,
+\,\oint_{\partial\Gamma}\chi_n^2(\lambda)\,
\delta\big(\mu_{\scriptscriptstyle\hspace{-.05cm}B}b^2
e^{\varphi_{\scriptscriptstyle\hspace{-.03cm}B}/2}\big)\,d\lambda\;.
\end{equation}
At this point, exploiting the spectral representation of the Green
function, i.e.
\begin{equation}
g(\zeta,\zeta')\,=\,\sum_{n\,\geqslant\,1}\,
\frac{\chi_n(\zeta)\chi_n(\zeta')}{\mu_n}
\end{equation}
we get the variation
\begin{eqnarray}\label{variation logdet FZZ}
\delta\left(\log \big(\textrm{Det}\,D\,\big)^{-1/2}\right)&=&
-\,\frac{1}{2}\,\sum_{n\,\geqslant\,1}\frac{\delta\mu_n}{\mu_n}\\
\rule{0pt}{.8cm}&=&
-\,2\int_{\Gamma}g(\zeta,\zeta)\,\delta\big(\mu
b^2e^{\varphi_{\scriptscriptstyle\hspace{-.03cm}B}}\big)\,
d^2\zeta\,-\,\frac{1}{2}\,\oint_{\partial\Gamma}
g_{\scriptscriptstyle\hspace{-.05cm}B}(\lambda,\lambda)\,
\delta\big(\mu_{\scriptscriptstyle\hspace{-.05cm}B}b^2
e^{\varphi_{\scriptscriptstyle\hspace{-.03cm}B}/2}\big)\,
d\lambda\nonumber
\end{eqnarray}
where the Green function at coincident points in the bulk and on
the boundary appear. Such quantities are divergent and have
to be regularized. \\
We have already learnt that the correct regularization in the bulk is
the one
suggested by Zamolodchikov and Zamolodchikov
\cite{ZZpseudosphere,MTpseudosphere}, i.e.
\begin{equation}\label{ZZ regularization bulk}
    g(\zeta,\zeta)\,\equiv\,\lim_{\zeta'\rightarrow\,
    \zeta}\,\left\{\,g(\zeta,\zeta')+\frac{1}{2}\,
\log\left|\,\zeta-\zeta'\right|^2\,\right\}
\end{equation}
while $g_{\scriptscriptstyle\hspace{-.05cm}B}(\lambda,\lambda)$
will be similarly defined by simply subtracting the logarithmic
divergence.\\
Notice that
$g_{\scriptscriptstyle\hspace{-.05cm}B}(\lambda,\lambda')$
diverges like $\log|\lambda-\lambda'|^2$ when $\lambda'\rightarrow
\lambda$ and not like $1/2\log|\lambda-\lambda'|^2$, as one could
naively expect. A general argument for this behavior is the
following\footnote{We are grateful to Giovanni Morchio for
providing the described argument.}.\\
After having transformed the simply connected domain $\Gamma$ into
the upper half plane $\mathbb{H}$, the Green function
$g_{\scriptscriptstyle\hspace{-.05cm}N}(\xi,\xi')$ for the
operator $D$ with Neumann boundary conditions satisfies
\begin{equation}\label{Neumann b.c. propagator}
\left(\,\frac{d}{d\xi}\,-\, \frac{d}{d\bar{\xi}}\,\right)
g_{\scriptscriptstyle\hspace{-.05cm}N}(\xi,\xi') \,=\,0\qquad
\hspace{.1cm}\textrm{when}\qquad \xi\in\mathbb{R}
\end{equation}
hence its behavior near the boundary ($\textrm{Im}\xi\rightarrow
0$) is given by the method of the images, i.e.
\begin{equation}
g_{\scriptscriptstyle\hspace{-.05cm}N}(\xi,\xi')\,=\,
-\,\frac{1}{2}\,\log(\xi-\xi')(\bar{\xi}-\bar{\xi}')
-\frac{1}{2}\,\log(\xi-\bar{\xi}')(\bar{\xi}-\xi')+\dots
\end{equation}
which satisfies (\ref{Neumann b.c. propagator}).\\
The complete Green function $g(\xi,\xi')$ with mixed boundary
conditions (\ref{FZZ b.c. propagator Gamma}) has the form
\begin{equation}\label{g=gNf}
g(\xi,\xi')\,=\,A(\xi,\xi')
\left(-\,\frac{1}{2}\,\log(\xi-\xi')(\bar{\xi}-\bar{\xi}')
-\frac{1}{2}\,\log(\xi-\bar{\xi}')(\bar{\xi}-\xi')+C(\xi,\xi')\right)
\end{equation}
where $A(\xi,\xi')$ and $C(\xi,\xi')$ are regular functions
\cite{Garabedian} with $A(\xi,\xi)=1$. The mixed boundary
conditions (\ref{FZZ b.c. propagator Gamma}) for $g(\xi,\xi')$
then read
\begin{equation}
\frac{g(\xi,\xi')}{A(\xi,\xi')}\left(\frac{d}{d\xi}-
\frac{d}{d\bar{\xi}}\right)
A(\xi,\xi')\,+\,A(\xi,\xi')\left(\frac{d}{d\xi}-
\frac{d}{d\bar{\xi}}\right) C(\xi,\xi')\,=\,-\,2\pi i\,
\mu_{\scriptscriptstyle\hspace{-.05cm}B}b^2\,
e^{\varphi_{\scriptscriptstyle\hspace{-.02cm}B}/2}\,g(\xi,\xi')
\end{equation}
for $\xi\in\mathbb{R}$, i.e.
\begin{equation}
\left\{\,\begin{array}{lcl} \displaystyle
\left(\frac{d}{d\xi}-\frac{d}{d\bar{\xi}}\right)
A(\xi,\xi')\,=\,-\,2\pi i\,
\mu_{\scriptscriptstyle\hspace{-.05cm}B}b^2\,
e^{\varphi_{\scriptscriptstyle\hspace{-.02cm}B}/2} A(\xi,\xi')&
\hspace{.5cm}\textrm{when}\hspace{.5cm}& \xi\in\mathbb{R}\\
\rule{0pt}{1cm}\displaystyle
\left(\frac{d}{d\xi}-\frac{d}{d\bar{\xi}}\right) C(\xi,\xi')\,=\,0
&\hspace{.5cm}\textrm{when}\hspace{.5cm}& \xi\in\mathbb{R}\;.
\end{array}\right.
\end{equation}
In the bulk for $\xi=\xi'$ we have
\begin{equation}\label{bulk divergence}
g(\xi,\xi')\,\simeq\,-\,\frac{1}{2}\,\log|\xi-\xi'|^2-\frac{1}{2}\,
\log|\xi-\bar{\xi}|^2+C(\xi,\xi)
\end{equation}
while for both $\xi=x$ and $\xi'=x'$ on the boundary
$\mathbb{R}=\partial\mathbb{H}$ we have
\begin{equation}
g(x,x')\,\simeq\,-\,\log|x-x'|^2+C(x,x)\;.
\end{equation}
We notice that the finite part of $g(\xi,\xi')$ in the bulk for
$\xi$ going to the boundary coincides with the finite part of
$g(x,x')$ on the boundary, which is given by $C(x,x)$. Such a
boundary behavior will be verified explicitly for the Green
function on the background generated by one source in section
\ref{Green function sect}, where also the finite terms at
coincident points will be
computed.\\
Thus, coming back to the general simply connected domain $\Gamma$,
we define the regularized value of the Green function on the
boundary at coincident points as follows
\begin{equation}\label{gB(lambda)}
g_{\scriptscriptstyle\hspace{-.05cm}B}(\lambda,\lambda)\,\equiv\,
\lim_{\lambda'\rightarrow\,\lambda}\,\left\{\,g(\lambda,\lambda')+
\log\left|\lambda-\lambda'\right|^2\,\right\}\;.
\end{equation}
Now we observe that, since the Green function $g(\zeta,\zeta')$ is
invariant in value under a conformal transformation
$\zeta\rightarrow \tilde{\zeta}=\tilde{\zeta}(\zeta)$, then its
regularized values at coincident points change as follows
\begin{equation}
g(\zeta,\zeta)\;\;\longrightarrow\;\;
\tilde{g}(\tilde{\zeta},\tilde{\zeta})\,=\,g(\zeta,\zeta)\,+\,
\frac{1}{2}\,
\log \bigg| \frac{d \tilde{\zeta}}{d
\zeta}\bigg|^{2}\hspace{1cm}\textrm{when}\hspace{1cm}\zeta\in\Gamma
\end{equation}
and
\begin{equation}
\hspace{1.15cm}
g_{\scriptscriptstyle\hspace{-.05cm}B}(\lambda,\lambda)
\;\;\longrightarrow\;\;
\tilde{g}_{\scriptscriptstyle\hspace{-.05cm}B}(\tilde{\lambda},
\tilde{\lambda})\,=\,
g_{\scriptscriptstyle\hspace{-.05cm}B}(\lambda,\lambda)\,+\, \log
\bigg| \frac{d \tilde{\zeta}}{d
\zeta}\bigg|^{2}\hspace{1,1cm},\hspace{1,1cm}
\zeta(\lambda)\in\partial\Gamma\;.
\end{equation}
We shall compute the change of (\ref{I1 def}) $I_1\rightarrow
\tilde{I}_1$ under a conformal transformation by computing the
transformation
properties of its derivatives w.r.t. $\eta_1,\dots,\eta_N$, $A$ and $l$.\\
The logarithmic variation of $\tilde{I}_1$ is given by
\begin{eqnarray}\label{variation tildeI1}
\delta\,\log\tilde{I}_1&=&
\delta\left(-\frac{1}{4\pi}\,\oint_{\partial\widetilde{\Gamma}}\tilde{k}\;
\tilde{\varphi}_{\scriptscriptstyle\hspace{-.05cm}B}\,
d\tilde{\lambda}\right)\\
\rule{0pt}{.8cm}& &
-\,2\int_{\widetilde{\Gamma}}\tilde{g}(\tilde{\zeta},\tilde{\zeta})\,
\delta\big(\mu
b^2e^{\tilde{\varphi}_{\scriptscriptstyle\hspace{-.03cm}B}}\big)\,d^2
\tilde{\zeta}
\,-\,\frac{1}{2}\,\oint_{\partial\widetilde{\Gamma}}
\tilde{g}_{\scriptscriptstyle\hspace{-.05cm}B}(\tilde{\lambda},
\tilde{\lambda})\,
\delta\big(\mu_{\scriptscriptstyle\hspace{-.05cm}B}b^2
e^{\tilde{\varphi}_{\scriptscriptstyle\hspace{-.03cm}B}/2}\big)\,
d\tilde{\lambda}\;.\nonumber
\end{eqnarray}
The terms in the second line can be rewritten as
\begin{eqnarray}\label{variation 1}
-\,2\int_{\Gamma}\tilde{g}(\tilde{\zeta},\tilde{\zeta})\,\delta\big(\mu
b^2e^{\varphi_{\scriptscriptstyle\hspace{-.03cm}B}}\big)\,d^2\zeta
\,-\,\frac{1}{2}\,\oint_{\partial\Gamma}
\tilde{g}_{\scriptscriptstyle\hspace{-.05cm}B}(\tilde{\lambda},
\tilde{\lambda})\,
\delta\big(\mu_{\scriptscriptstyle\hspace{-.05cm}B}b^2
e^{\varphi_{\scriptscriptstyle\hspace{-.03cm}B}/2}\big)\,d\lambda
&=&\\
\rule{0pt}{.8cm}& &\hspace{-7.4cm}
=\;-\,2\int_{\Gamma}g(\zeta,\zeta)\,\delta\big(\mu
b^2e^{\varphi_{\scriptscriptstyle\hspace{-.03cm}B}}\big)\,d^2\zeta
\,-\,\frac{1}{2}\,\oint_{\partial\Gamma}
g_{\scriptscriptstyle\hspace{-.05cm}B}(\lambda,\lambda)\,
\delta\big(\mu_{\scriptscriptstyle\hspace{-.05cm}B}b^2
e^{\varphi_{\scriptscriptstyle\hspace{-.03cm}B}/2}\big)\,
d\lambda\nonumber\\
\rule{0pt}{.8cm}& &\hspace{-6.8cm} -\int_{\Gamma}\log
J\;\delta\big(\mu
b^2e^{\varphi_{\scriptscriptstyle\hspace{-.03cm}B}}\big)\,
d^2\zeta
\,-\,\frac{1}{2}\,\oint_{\partial\Gamma} \log J\;
\delta\big(\mu_{\scriptscriptstyle\hspace{-.05cm}B}b^2
e^{\varphi_{\scriptscriptstyle\hspace{-.03cm}B}/2}\big)\,
d\lambda\nonumber
\end{eqnarray}
where $J\equiv|d\tilde{\zeta}/d\zeta|^2$ is independent of
$\eta_1,\dots,\eta_N$, $A$ and $l$.\\
Using the Liouville equation (\ref{liouville eq with sources
varphi BLFT}) for $\varphi_{\scriptscriptstyle\hspace{-.05cm}B}$
and the boundary conditions (\ref{FZZ b.c. classic field}), we
obtain for last two terms in (\ref{variation 1})
\begin{equation}
-\,\delta\left(\,\sum_{j\,=\,1}^{N}\eta_j\left.\log
J\right|_{\,\zeta_j}\right)+\,\delta\left[\,\frac{1}{8\pi
i}\oint_{\partial\Gamma}\hspace{-.1cm}
\varphi_{\scriptscriptstyle\hspace{-.05cm}B}\Big(\partial_\zeta
\log J\,d\zeta \,-\,\partial_{\bar{\zeta}} \log
J\,d\bar{\zeta}\Big)+
\frac{1}{4\pi}\oint_{\partial\Gamma}\hspace{-.1cm} k\,\log
J\,d\lambda\,\right].
\end{equation}
The term in (\ref{variation tildeI1}) containing the curvature
$\tilde{k}$ becomes
\begin{equation}
\delta\left[\,-\,\frac{1}{4\pi}\,\oint_{\partial\Gamma}k\;
\varphi_{\scriptscriptstyle\hspace{-.05cm}B}\,d\lambda\,-\,
\frac{1}{8\pi
i}\oint_{\partial\Gamma}\hspace{-.1cm}
\varphi_{\scriptscriptstyle\hspace{-.05cm}B}\Big(\partial_\zeta
\log J\,d\zeta \,-\,\partial_{\bar{\zeta}} \log
J\,d\bar{\zeta}\Big)-\,
\frac{1}{4\pi}\oint_{\partial\widetilde{\Gamma}}\hspace{-.1cm}
\tilde{k}\,\log \tilde{J}\,d\tilde{\lambda}\,\right]
\end{equation}
where we have used the transformation law (\ref{curvature
transformation}) for $k$ under conformal transformations.\\
Summing the two contributions and taking into account that the
term
\begin{equation}
\frac{1}{4\pi}\left(\,
\oint_{\partial\widetilde{\Gamma}}\tilde{k}\,\log\tilde{J}
\,d\tilde{\lambda}\,-\,\oint_{\partial\Gamma}k\,\log J
\,d\lambda\right)
\end{equation}
does not depend on $\eta_1,\dots,\eta_N$, $A$ and $l$,  we find
that
\begin{equation}
\delta\,\log\tilde{I}_1\,=\,\delta\,\log
I_1\,-\,\delta\left(\,\sum_{j\,=\,1}^{N}\eta_j\left.\log
J\right|_{\,\zeta_j}\right)
\end{equation}
which gives
\begin{equation}\label{variation I1}
\log\tilde{I}_1\,=\,\log
I_1\,-\,\sum_{j\,=\,1}^{N}\eta_j\left.\log
J\right|_{\,\zeta_j}+\,f(\zeta_1,\dots,\zeta_N)
\end{equation}
where $f(\zeta_1,\dots,\zeta_N)$ is independent of
$\eta_1,\dots,\eta_N$, $A$ and $l$. Since for vanishing $\eta_1$
the vertex correlation function has to be independent of
$\zeta_1$, we have that $f(\zeta_1,\dots,\zeta_N)$ does not depend
on $\zeta_1$ and, similarly, on $\zeta_2,\dots,\zeta_N$.\\
As the conformal dimensions $\Delta_{\alpha_k}$ are given by
\begin{equation}
-\,\Delta_{\alpha_k}\frac{\partial\log
J|_{\,\zeta_k}}{\partial\zeta_k}\,=\,\frac{\partial}{\partial\zeta_k}\,
\log\frac{\langle
\, e^{2\alpha_1\tilde{\phi}(\tilde{\zeta}_1)}\dots
e^{2\alpha_N\tilde{\phi}(\tilde{\zeta}_N)} \,\rangle}{\langle \,
e^{2\alpha_1\phi(\zeta_1)}\dots e^{2\alpha_N\phi(\zeta_N)}
\,\rangle}
\end{equation}
the relation (\ref{variation I1}) provides the one loop quantum
correction to the semiclassical dimensions
\begin{equation}
\frac{\eta(1-\eta)}{b^2}\hspace{.8cm}\longrightarrow\hspace{.8cm}
\Delta_{\eta/b}\,=\,\frac{\eta(1-\eta)}{b^2}\,+\,\eta\,=\,
\alpha\left(\,\frac{1}{b}+b-\alpha\right)
\end{equation}
which coincide with the exact quantum dimensions \cite{CT}. In
particular the weights of the bulk cosmological term $e^{2b\phi}$
become $(1,1)$.

\section{The one point function}
\label{one point section}

\noindent Through a conformal transformation, one can always
reduce the finite simply connected domain $\Gamma$ to the unit
disk $\Delta$. The classical and the quantum actions are given by
(\ref{action classical phiB BLFT}) and (\ref{quantumaction BLFT
3}) respectively, with $k=1$. The parametric boundary length in
the case of the unit disk $\Delta$ is given by the angular
coordinate
$\theta$.\\
We shall consider the one point function, i.e. one single source
of charge $\eta_1=\eta$ placed in $z_1=0$, without loss of
generality.

\subsection{The classical action}

The solution of the Liouville equation (\ref{liouville eq with
sources varphi BLFT}) with $N=1$ on the unit disk is \cite{FZZ,
Moore Seiberg Staudacher}
\begin{equation}\label{phiclassic BLFT}
    e^{\varphi_{c}}\,=\,\frac{1}{\pi\mu b^2}\;
    \frac{a^2(1-2\eta)^2}{\big((z\bar{z})^{\eta}
    -a^2(z\bar{z})^{1-\eta}\big)^2}
    \hspace{2cm}
    \mu>0\,,\hspace{.5cm}\;0<a^2 < 1
\end{equation}
with $\mu > 0$ and $1-2\eta > 0$. The condition $a^2 < 1$ is
necessary to avoid singularities inside $\Delta$ except for the
one placed in $0$. The boundary conditions (\ref{FZZ b.c. classic
field}) when $\Gamma=\Delta$ read
\begin{equation}\label{FZZ b.c. classic Delta}
-\,r^2\partial_{r^2}\varphi_{c}\,=\,1\,+\,2\pi\,
\mu_{\scriptscriptstyle\hspace{-.05cm}B}b^2\,e^{\varphi_{c}/2}
\qquad\hspace{.3cm}\textrm{when}\hspace{.3cm}\qquad r\equiv |z|=1
\end{equation}
and this condition on the solution (\ref{phiclassic BLFT})
provides the following relation between $a^2$ and the scale
invariant ratio of the cosmological constants
\begin{equation}\label{FZZ b.c. classic a^2}
    \sqrt{\pi}\,b\,
    \frac{\mu_{\scriptscriptstyle\hspace{-.05cm}B}}{\sqrt{\mu}}\,=\,
    -\,\frac{1+a^2}{2|a|}\;.
\end{equation}
It is important to remark that the semiclassical limit can be
realized only for $\mu_{\scriptscriptstyle\hspace{-.05cm}B}<0$.
More precisely, from (\ref{FZZ b.c. classic a^2}), we find that
the scale invariant ratio of the cosmological constants has to be
$\sqrt{\pi}\,b\,\mu_{\scriptscriptstyle\hspace{-.05cm}B}/\sqrt{\mu}<-1$.\\
The classical field (\ref{phiclassic BLFT}) gives rise to specific
expressions for the area $A$ and the boundary length $l$ of the
unit disk in terms of the bulk cosmological constant $\mu$, the
charge $\eta$ and parameter $a^2$
\begin{eqnarray}
   A \hspace{-.1cm} &=& \hspace{-.2cm}\int_{\Delta} e^{\varphi_{c}}
    d^2z\;\,=\;\frac{1}{\mu b^2}\;\frac{a^2(1-2\eta)}{1-a^2}\\
    \label{length classic}
\rule{0pt}{1cm}l \hspace{-.1cm}& = & \hspace{-.2cm}
\oint_{\partial\Delta} e^{\varphi_{c}/\,2}\,
    d\theta\;=\;\frac{\sqrt{\pi}}{b\,\sqrt{\mu}}
    \;\frac{2 |a|\,(1-2\eta)}{1-a^2}\;=\;-\,
    \frac{1}{\mu_{\scriptscriptstyle\hspace{-.05cm}B}b^2}\;
    \frac{(1-2\eta)(1+a^2)}{1-a^2}
\end{eqnarray}
where in the last step of (\ref{length classic}) we have employed
(\ref{FZZ b.c. classic a^2}). A useful relation we shall employ in
the following is
\begin{equation}\label{a^2(A,l)}
a^2 \,=\, 1-\,4\pi\,\frac{A}{l^2}\,(1-2\eta)\;.
\end{equation}
Given the classical solution (\ref{phiclassic BLFT}), we can
compute the classical action (\ref{action classical phiB BLFT}) on
such a background. The result is
\begin{equation}\label{classical action area length}
S_{cl}[ \,\varphi_{c}\,] \,=\,
\frac{S_{0}(\eta;A,l)}{b^2}\,+\mu\,A+
\mu_{\scriptscriptstyle\hspace{-.05cm}B}\,l
\end{equation}
where \cite{FZZ}
\begin{eqnarray}\label{FZZ classic action A l}
S_{0}(\eta;A,l)\,=\, \left.\rule{0pt}{.4cm}b^2\,S_{cl}[
\,\varphi_{c}\,]
\right|_{\,\mu\,=\,\mu_{\scriptscriptstyle\hspace{-.02cm}B}\,=\,0}
& & \hspace{-.5cm}=\;\frac{l^2}{4\pi
\,A}\,+(1-2\eta)\left(\,\log\frac{2A}{l}\,+\log(1-2\eta)-1\,\right)
\nonumber\\
\rule{0pt}{.8cm}\label{FZZ classic action mu muB} &
&\hspace{-3.2cm}\;=\;
(1-2\eta)\left(\,\frac{1}{1-a^2}+\log|a|-\,\frac{1}{2}\,\log(\pi\mu
b^2)+\log(1-2\eta)-1\right)\;.\nonumber\\
& &
\end{eqnarray}

\subsection{The Green function}
\label{Green function sect}

\noindent The Green function on the background generated by one
heavy charge satisfies the following equation
\begin{eqnarray}\label{g(z,z') equation FZZ}
    D\,g(z,t\hspace{.03cm})&=&
    \left(-\,\frac{2}{\pi}\,\partial_z\partial_{\bar{z}}\,+ \,4\,\mu
    b^2\,e^{\varphi_{c}}\right)g(z,t\hspace{.03cm}) \\
\rule{0pt}{.8cm}
&=&\left(-\,\frac{2}{\pi}\,\partial_z\partial_{\bar{z}}\,+ \,
\frac{4\,a^2(1-2\eta)^2}{\pi\big((z\bar{z})^{\eta}
    -a^2(z\bar{z})^{1-\eta}\big)^2}\right)g(z,t\hspace{.03cm})
    \;\;=\;\;\delta^2(z-t\hspace{.02cm})\nonumber
\end{eqnarray}
and its boundary conditions are
\begin{equation}\label{FZZ b.c. propagator}
-\,r^2\frac{\partial}{\partial r^2}\,g(z,t\hspace{.02cm})\,=\,\pi
\,\mu_{\scriptscriptstyle\hspace{-.05cm}B}b^2\,e^{\varphi_{c}/2}\,
g(z,t\hspace{.03cm})
\qquad\hspace{.3cm}\textrm{when}\hspace{.3cm}\qquad r^2=1
\end{equation}
where $z=r e^{i\theta}$ and $\varphi_{c}$ is the classical
background field (\ref{phiclassic BLFT}). Exploiting the relation
(\ref{FZZ b.c. classic a^2}) derived from the boundary conditions
of $\varphi_{c}$, the boundary conditions for the Green function
read
\begin{equation}\label{FZZ b.c. propagator complex}
\big(\,z\,\partial_z+\bar{z}\,\partial_{\bar{z}}\,\big)\,g(z,t)
\,=\,(1-2\eta)\,\frac{1+a^2}{1-a^2}\;g(z,t)
\qquad\hspace{.3cm}\textrm{when}\hspace{.3cm}\qquad |z|=1\;.
\end{equation}

\noindent To compute $g(z,t)$ in the simplest way, we expand it as
a sum of partial waves
\begin{equation}\label{Fourier sum FZZ}
   g(z,t\hspace{.02cm})
   \,=\,\sum_{m\,\geqslant\,0}\,g_m(x,y)\,\cos\big(m(\theta_x-\theta_y)\big)
\end{equation}
where $x=|z|^2$ and $y=|t|^2$. The Fourier coefficients $g_m(x,y)$
are symmetric in the arguments and satisfy the following equation
\begin{equation}\label{eq for gm}
    \left(-\,2\,\frac{\partial}{\partial x}\left(x\,\frac{\partial}
{\partial x}\,\right)
    +\,\frac{m^2}{2\,x}+\,\frac{4\,a^2(1-2\eta)^2}{(x^\eta-a^2
    x^{1-\eta})^2}\,\right) g_m(x,y)\,=\, d_m
    \,\delta(x-y)
\end{equation}
with $d_0=1$ and $d_{m}=2$ for $m\geqslant 1$. They are given by
\begin{equation}
   g_m(x,y)\,=\,\theta(y-x)\,a_m(x)\,b_m(y) +
   \theta(x-y)\,a_m(y)\,b_m(x)\hspace{0.5cm},
\hspace{0.5cm}m \geqslant0
\end{equation}
where both $a_m(x)$ and $b_m(x)$ satisfy the homogenous version of
(\ref{eq for gm}). The functions $a_m(x)$ must be regular in $x=0$
and, to reproduce the delta singularity, the wronskian of the
solutions $a_m$ and $b_m$ must be
\begin{equation}\label{wronskian}
    \left\{\,\begin{array}{l}
\displaystyle \frac{\partial a_0(r^2)}{\partial r}\,b_0(r^2)\,-\,
\frac{\partial
b_0(r^2)}{\partial r}\,a_0(r^2) \;=\; \frac{1}{r}\hspace{.5cm}\\
\displaystyle \rule{0pt}{.9cm} \frac{\partial a_m(r^2)}{\partial
r}\,b_m(r^2)\,-\, \frac{\partial b_m(r^2)}{\partial r}\,a_m(r^2)
\;=\;\frac{2}{r}\hspace{0.7cm},\hspace{1.0cm}m \geqslant1\;.
             \end{array}
    \right.
\end{equation}
The boundary conditions (\ref{FZZ b.c. propagator}) are translated
into
\begin{equation}
2\,y\,\frac{\partial}{\partial
y}\,b_m(y)\,=\,(1-2\eta)\,\frac{1+a^2}{1-a^2}\,
\rule{0pt}{.4cm}b_m(y)
\qquad\hspace{.3cm}\textrm{when}\hspace{.3cm}\qquad y\,=\,1\hspace{0.5cm},
\hspace{0.5cm}m \geqslant0\;.
\end{equation}
The solutions for $m=0$ are
\begin{equation}\label{am functions FZZ}
\rule{0pt}{.7cm} a_0(x) \,=\, \frac{1+a^2 x^{1-2\eta}}{1-a^2
x^{1-2\eta}} \hspace{1.4cm} b_0(y) \,=\, -\,\frac{1}{2(1-2\eta)}\,
\left(\,\frac{1+a^2y^{1-2\eta}}{1-a^2y^{1-2\eta}}\,\log
y^{1-2\eta} + 2\,\right)
\end{equation}
while $a_m(x)$ and $b_m(y)$ for $m\geqslant 1$ read
\begin{eqnarray}
\rule{0pt}{.6cm} a_m(x) \hspace{-.1cm}& = &\hspace{-.1cm}
\frac{x^{m/2}}{1-a^2x^{1-2\eta}}\,\left(\,1-\frac{m-(1-2\eta)}
{m+(1-2\eta)}\;a^2x^{1-2\eta}\,\right) \\
\rule{0pt}{1cm} b_m(y)\hspace{-.1cm} & = &\hspace{-.1cm}
-\,\frac{y^{-m/2}}{m\big(m-(1-2\eta)\big)}\,\left(\,(1-2\eta)\,
\frac{1+a^2y^{1-2\eta}}{1-a^2y^{1-2\eta}}
\,(1-y^m) - m (1+y^m)\,\right)\;.\hspace{1cm}
\end{eqnarray}
For $a^2\rightarrow 1$, the expressions of $a_m(x)$ and $b_m(y)$
go over to their counterparts on the pseudosphere
\cite{MTpseudosphere}.\\
Given $a_m(x)$ and $b_m(y)$, the series (\ref{Fourier sum FZZ})
can be explicitly summed \cite{MTpseudosphere,Bateman}. The result
is
\begin{eqnarray}\label{g(z,t) FZZ final form}
 g(z,t) & = & -\,\frac{1}{\rule{0pt}{.4cm}2}\;
\frac{1+a^2(z\bar{z})^{1-2\eta}}{\rule{0pt}{.4cm}1-a^2(z\bar{z})^{1-2\eta}}
\;\left\{\;
\frac{1+a^2(t\bar{t}\hspace{.04cm})^{1-2\eta}}{
\rule{0pt}{.4cm}1-a^2(t\bar{t}\hspace{.04cm})^{1-2\eta}}
\;\log\omega(z,t) +\,\frac{2}{\rule{0pt}{.4cm}1-2\eta}\;\right\}
\\
\rule{0pt}{.8cm} & & -\,
\frac{1}{\rule{0pt}{.4cm}\,1-a^2(z\bar{z})^{1-2\eta}}\;
\frac{1}{\rule{0pt}{.4cm}\,1-a^2(t\bar{t}\hspace{.04cm})^{1-2\eta}}\,
\times
\nonumber \\
\rule{0pt}{1cm} & & \hspace{-2.5cm}\times \,
\left\{\;a^2\,\frac{(t\bar{t}\hspace{.04cm})^{1-2\eta}}{
\rule{0pt}{.4cm}2\eta}\;
\frac{z}{t}\;F(\,2\eta,1;1+2\eta
;\,z/t\,)+\,a^2\,\frac{(z\bar{z})^{1-2\eta}}{\rule{0pt}{.4cm}2(1-\eta)}\;
\frac{z}{t}\;F(\,2-2\eta,1;3-2\eta;\,z/t\,)\,+\,\textrm{c.c.}\;\right.
\nonumber \\
\rule{0pt}{.9cm} & &\hspace{-2cm}\left.
-\,\frac{1}{\rule{0pt}{.4cm}2\eta}\;z\,\bar{t}\;
F(\,2\eta,1;1+2\eta;\,z\,\bar{t}\,)\,-\,
a^4\,\frac{(z\bar{z})^{1-2\eta}(t\bar{t}\hspace{.04cm})^{1-2\eta}}{
\rule{0pt}{.4cm}2(1-\eta)}
\;z\,\bar{t}\;F(\,2-2\eta,1;3-2\eta;\,z\,\bar{t}\,)\,+\,\textrm{c.c.}\,
\right\}.
\nonumber
\end{eqnarray}
This Green function can be also obtained by applying the general
method developed in \cite{MTpseudosphere,MTheavyZZ,MV}.\\
In the limit $a^2\rightarrow 1$ for $z$ and $t$ fixed we recover
the Green function of the pseudosphere
\cite{MTpseudosphere,MTheavyZZ}, which has also a well defined
limit $\eta\rightarrow 0$. On the other hand the limit
$\eta\rightarrow 0$ of $g(z,t)$ for fixed $a^2<1$ is singular and
this fact is related to the occurrence of a zero mode when
$\eta=0$ (see appendix \ref{spectrum}). Thus the two limits
$a^2\rightarrow 1$ and $\eta\rightarrow 0$ of the Green function
(\ref{g(z,t) FZZ final form}) do not commute.\\

\noindent The regularized value $g(z,z)$ of this Green function at
coincident point is defined in (\ref{ZZ regularization bulk}). To
compute it, we can expand $\log|z-t|^2$ as a Fourier series with
symmetric and factorized coefficients by employing
\begin{equation}\label{Fourier sum log}
  \frac{1}{2}\, \log|z-t|^2
   \,=\,\frac{1}{2}\,\log y\,-\sum_{m\,\geqslant\,1}\,
   \frac{1}{m}\left(\frac{x}{y}\right)^{m/2}
   \cos\big(m(\theta_x-\theta_y)\big)
\end{equation}
where $x=\textrm{min}(|z|,|t|)$ and $y=\textrm{max}(|z|,|t|)$.
Adding (\ref{Fourier sum log}) to (\ref{Fourier sum FZZ}) and
computing the result at coincident points, we get a series
representation for $g(z,z)$, which can be summed explicitly.\\
Otherwise, we can apply directly the definition (\ref{ZZ
regularization bulk}) to (\ref{g(z,t) FZZ final form}), obtaining
the same result, i.e.
\begin{eqnarray}\label{g(z,z) FZZ}
\hspace{-0cm} g(z,z) \hspace{-.1cm}& = &\hspace{-.1cm}
\Bigg(\,\frac{1+a^2(z\bar{z})^{1-2\eta}}{\rule{0pt}
{.4cm}1-a^2(z\bar{z})^{1-2\eta}}\,\Bigg)^2 \log(1-z\bar{z})\,
-\,\frac{1}{1-2\eta}\;\frac{1+a^2(z\bar{z})^{1-2\eta}}
{\rule{0pt}{.4cm}1-a^2(z\bar{z})^{1-2\eta}}
\\
\rule{0pt}{1cm}
 & &   +\,\frac{2\,(z\bar{z})^{1-2\eta}}{
\big(\rule{0pt}{.45cm}\,1-a^2(z\bar{z})^{1-2\eta}\,\big)^2}\,
\left( B_{z\bar{z}}\big(2\eta\,,0\big)+a^4
B_{z\bar{z}}\big(2-2\eta\,,0\big)
\rule{0pt}{.5cm}\right.\nonumber\\
& &\hspace{5.6cm} \left.\rule{0pt}{.5cm}+a^2\big(\,
2\gamma_{\scriptscriptstyle\hspace{-.02cm}E}+
\psi(2\eta)+\psi(2-2\eta)-\log
z\bar{z}\,\big) \right)\;.\nonumber
\end{eqnarray}
where $\gamma_{\scriptscriptstyle\hspace{-.02cm}E}$ is the Euler
constant and $\psi(x)=\Gamma'(x)/\Gamma(x)$.\\
For $a^2\rightarrow 1$ we have that $g(z,z)$ in the bulk becomes the
corresponding function on the pseudosphere
\cite{MTpseudosphere,MTheavyZZ}, hence the two limits
$a^2\rightarrow 1$ and $t\rightarrow z$ of the Green function
(\ref{g(z,t) FZZ final form}) commute.\\
By using the expansion of the incomplete Beta function
$B_x(\alpha,0)$ around $x=1$ \cite{MTpseudosphere,Bateman}, we
find that the boundary behavior of $g(z,z)$ is
\begin{eqnarray}\label{g(z,z) at boundary}
g(z,z)&=&-\,\log(1-z\bar{z})\,-\,\frac{1}{1-2\eta}\,-2\,
\gamma_{\scriptscriptstyle\hspace{-.02cm}E}
-\,2\,\psi(1-2\eta)\,+\,\frac{2\pi
\cot(2\pi\eta)}{1-a^2}\,\nonumber\\
\rule{0pt}{.7cm}& & +\;O\big((1-z\bar{z})\log(1-z\bar{z})\big)\;.
\end{eqnarray}
We notice from this formula that the two limits $a^2\rightarrow 1$
and $|z|\rightarrow 1$ of $g(z,z)$ do not commute.\\
The regularized value of the Green function on the boundary is
defined in (\ref{gB(lambda)}). Again, its explicit expression can
be obtained either by taking the limit (\ref{gB(lambda)}) on
(\ref{g(z,t) FZZ final form}) or by summing explicitly the series
given by
\begin{equation}
g\big(e^{i\theta},e^{i\theta'}\big)\,=\,a_0(1)\,b_0(1)+
\sum_{m\,\geqslant\,1}
a_m(1)\,b_m(1)\,\cos\big(m(\theta-\theta')\big)
\end{equation}
and
\begin{eqnarray}
-\sum_{m\,\geqslant\,1}\,\frac{2}{m}\,\cos\big(m(\theta-\theta')\big)&=&
\log\big|\,e^{i\theta}-e^{i\theta'}\big|^2\;\;=\;\;
\log\big(2-2\cos(\theta-\theta')\big)\nonumber\\
&=&2\,\log\left|\theta-\theta'\right|+O\big((\theta-\theta')^2\big)\;.
\end{eqnarray}
The result is
\begin{equation}\label{g(theta)}
g_{\scriptscriptstyle\hspace{-.05cm}B}(\theta,\theta)\,=
\,-\,\frac{1}{1-2\eta}\,
-\,2\,\gamma_{\,\scriptscriptstyle\hspace{-.05cm}E}-
\,2\,\psi(1-2\eta)\,+\,\frac{2\pi\cot(2\pi\eta)}{1-a^2}
\end{equation}
which is independent of $\theta$ by rotational invariance.\\
We notice that
$g_{\scriptscriptstyle\hspace{-.05cm}B}(\theta,\theta)$ coincides
with the finite part of $g(z,z)$ when $|z|\rightarrow 1$, as shown
in general in section \ref{constrained pathint section}.

\subsection{Fixed area and boundary length expansion}

\noindent At the semiclassical level, formula (\ref{FZZ b.c.
classic a^2}) coming from the boundary conditions (\ref{FZZ b.c.
classic field}) for the classical field $\varphi_{c}$ tells us
that $\mu_{\scriptscriptstyle\hspace{-.05cm}B} \hspace{-.04cm}<
0$; hence, from (\ref{action bound Gamma}), we have to work at
least with fixed boundary length $l$. The semiclassical value of
the action at fixed area $A$ and fixed boundary length $l$ has
been computed in \cite{FZZ}
and it has been reported in (\ref{FZZ classic action A l}). \\
To compute the quantum determinant at fixed area and boundary
length, we perform a constrained functional integral by exploiting
the results obtained in section \ref{constrained pathint section}
for the $N$ point functions. For the one point function, (\ref{N
point geometric}) becomes
\begin{equation}\label{one point fixed A and l def}
\left\langle \, e^{2(\eta/b)\phi(0)}
\,\right\rangle\,\equiv\,U(\eta;\mu,
\mu_{\scriptscriptstyle\hspace{-.05cm}B})\,\equiv\,
\int_{0}^{\infty}\frac{dl}{l}\;
e^{-\mu_{\scriptscriptstyle\hspace{-.05cm}B}
l} \int_{0}^{\infty}\frac{dA}{A}\;e^{-\mu
A}\,Z(\eta;A,l\hspace{.04cm})\;.
\end{equation}
In order to understand the dependence of
$Z(\eta;A,l\hspace{.04cm})$ on its arguments, it is useful to
define $\hat{\varphi}_{c}$ as follows
\begin{equation}\label{phiclassic BLFT hat}
    e^{\varphi_{c}}\,=\,\left(\frac{l}{2\pi}\right)^2
    \frac{(1-a^2)^2}{\big((z\bar{z})^{\eta}
    -a^2(z\bar{z})^{1-\eta}\big)^2}\,\equiv\,
    \left(\frac{l}{2\pi}\right)^2 e^{\hat{\varphi}_{c}}
\end{equation}
where $\hat{\varphi}_{c}$ depends only on $\eta$ and $a^2$. Using
\begin{equation}
e^{-\frac{1}{4\pi}\oint_{\partial\Delta}
\varphi_{c}\,d\theta}\,=\,\frac{2\pi}{l}
\end{equation}
and the definition (\ref{phiclassic BLFT hat}) of
$\hat{\varphi}_{c}$, from (\ref{fixed A and l one loop}) we find
to one loop
\begin{equation}\label{Z(A,l) uno}
Z(\eta;A,l\hspace{.04cm})=
e^{-S_{0}(\eta;A,l)/b^2}\,\frac{(2\pi)^4A}{2\,b^2
l^3}\int\hspace{-.06cm}\mathcal{D}\,[\, \chi
\,]\;e^{-\frac{1}{2}(\chi,D\chi)} \;\delta\left(\int_{\Delta}
e^{\hat{\varphi}_{c}}
    \chi\,d^2z\right)
\delta\left(e^{\hat{\varphi}_{c}(1)/2}\oint_{\partial\Delta}
    \chi\;d\theta\right).
\end{equation}
Exploiting the relation (\ref{a^2(A,l)}), we
get the following structure
\begin{eqnarray}
Z(\eta;A,l\hspace{.04cm})&=&e^{-S_{0}(\eta;A,l)/b^2}\;
\frac{(2\pi)^4A}{2\,b^2
l^3}\;
f(\eta,A/l^2)\big(1+O(b^2)\big)\\
\rule{0pt}{.8cm}&=&
e^{-S_{0}(\eta;A,l)/b^2}\;\frac{(2\pi)^4A}{2\,b^2 l^3}\;
f_1(\eta,a^2)\big(1+O(b^2)\big)\;.\nonumber
\end{eqnarray}
After expanding $\chi(z)$ in circular harmonics
\begin{equation}
\chi(z)=\sum_{m\geqslant0}\chi_m(x)\cos(m\theta)\hspace{1cm},
\hspace{1cm}x=|z|^2
\end{equation}
we notice that the constraints involve only the $m=0$ component of
the quantum field $\chi(z)$; hence we are left with the following
constrained quadratic path integral to one loop
\begin{equation}\label{constrained func integral}
Z(\eta;A,l\hspace{.04cm})\,=\,
e^{-S_{0}(\eta;A,l)/b^2}\,\frac{(2\pi)^2 A}{2\,b^2
l^3}\int\hspace{-.06cm}\mathcal{D}\,[\, \chi
\,]\,e^{-\frac{1}{2}\,(\chi,D\chi)}\; \delta\left(\int_{0}^1
e^{\hat{\varphi}_{c}}
    \chi_0(x)\,dx\right)
\delta\Big(e^{\hat{\varphi}_{c}(1)/2}\,\chi_0(1)\Big).
\end{equation}
The integrations over the partial waves with $m\neq 0$ give no
problems because the constraints involve only the $m=0$ sector of
the quadratic functional integral (\ref{constrained func
integral}).

\subsection{The $m=0$ sector}
\label{m0 sector sec}

In this subsection we shall examine the $m=0$ subspace. In
appendix \ref{spectrum} is proved that the operator $D_0$, i.e.
$D$ acting on the $m=0$ subspace, has one and only one negative
eigenvalue. To simplify the notation, we shall denote by
$\zeta(z)$ the field $\chi_0(z)$, by $\zeta_1(z)$ the normalized
eigenfunction of $D_0$ associated to the unique eigenvalue
$\mu_1=(2/\pi)\lambda_1<0$ and by $\zeta_{\perp}(z)$ the component
of $\chi_0(z)$ orthogonal to $\zeta_1(z)$.\\
First we prove that the fixed boundary length constraint is
sufficient to make the functional integral (\ref{constrained func
integral}) stable. Exploiting the integral representation of the
$\delta$ function, the fixed boundary length constrained path
integral is given by
\begin{eqnarray}
Y \hspace{-.1cm}&=&\hspace{-.1cm}
\frac{1}{2\pi}\int\hspace{-.06cm}\mathcal{D}\,[\,\zeta\,]\, \int
d\tau\; \exp\left\{-\frac{1}{2}\,\big(\zeta,D_0\zeta\big)
+i\,\tau\,e^{\hat{\varphi}_{c}(1)/2}\,\zeta(1)\,\right\} \\
\rule{0pt}{.9cm}& &\hspace{-1.1cm}=\;
\frac{1}{2\pi}\int\hspace{-.06cm}\mathcal{D}\,[\,\zeta_{\perp}\,]
\int d\tau\int dc_1\,
\exp\left\{-\frac{\mu_1}{2}\,c_1^2-\frac{1}{2}\,\big(\zeta_{\perp},
D_0\zeta_{\perp}\big)
+i\,\tau\,e^{\hat{\varphi}_{c}(1)/2}\,\big(c_1\zeta_1(1)+
\zeta_{\perp}(1)\big)
\right\}\nonumber
\end{eqnarray}
where
$\zeta(z)=c_1\zeta_1(z)+\zeta_{\perp}(z)=\sum_{n=1}^{+\infty}c_n\zeta_n(z)$.
Now we perform the following change of variable
\begin{equation}
\zeta_{\perp}(z)\,=\, \zeta_{\perp}'(z)\,+\,i\,\tau
\,g_{0\perp}(z,1)\,e^{\hat{\varphi}_{c}(1)/2}
\end{equation}
where
\begin{equation}
g_{0\perp}(z,z')\,=\,
\sum_{n\,\geqslant\,2}\,\frac{\zeta_n(z)\zeta_n(z')}{\mu_n}
\end{equation}
is the Green function of the $m=0$ sector orthogonal to the mode
$\zeta_1(z)$. Then, integrating in $\tau$, we find
\begin{equation}\label{constained length integ}
Y\,=\,\frac{1}{\sqrt{2\pi\,e^{\hat{\varphi}_{c}(1)}g_{0\perp}(1,1)}}
\int\hspace{-.06cm}\mathcal{D}\,[\,\zeta_{\perp}'\,]\, \int dc_1\;
\exp\left\{-\frac{1}{2}\,\big(\zeta_{\perp}',D_0\zeta_{\perp}'\big)
-\frac{c_1^2}{2}\left(\mu_1+\frac{\zeta_1^2(1)}{g_{0\perp}(1,1)}\right)\,
\right\}
\end{equation}
where $g_{0\perp}(1,1)>0$ because $\mu_j>0$ for $j\geqslant 2$.
The coefficient of $-c_1^2/2$ can be written in the following form
\begin{equation}
\frac{\mu_1}{g_{0\perp}(1,1)}\;g_{0}(1,1)
\end{equation}
from which one immediately sees that it is strictly positive,
being
\begin{equation}
g_0(1,1)\,=\,a_0(1)\,b_0(1)\,=\,-\,\frac{1+a^2}{(1-2\eta)(1-a^2)}\,<\,0\;.
\end{equation}
Now we can integrate in $c_1$ and the final result for $Y$ is
\begin{equation}
Y\,=\,\frac{1}{\sqrt{-\mu_1}\sqrt{-e^{\hat{\varphi}_{c}(1)}g_{0}(1,1)}}\,
\int\hspace{-.06cm}\mathcal{D}\,[\,\zeta_{\perp}\,]\;
e^{-\frac{1}{2}(\zeta_{\perp},D\zeta_{\perp})}\;.
\end{equation}
This procedure shows that in spite of
$\mu_1<0$ the constrained integral is stable.\\

\noindent Thus one could work keeping fixed $\mu$ and the boundary
length $l$. Instead, to compare our results with the ones obtained
in \cite{FZZ}, we introduce also the fixed area constraint.
Exploiting again the integral representation of the $\delta$
functions, the functional integral for the $m=0$ wave coming from
(\ref{constrained func integral}) reads
\begin{equation}
\frac{1}{(2\pi)^2}\int\hspace{-.06cm}\mathcal{D}\,[\,\zeta\,]\,
\int d\rho\int d\tau\;
\exp\left\{-\frac{1}{2}\,\big(\zeta,D_0\zeta\big) +i\,\rho\int_0^1
e^{\hat{\varphi}_{c}}\zeta(x)\,dx
+i\,\tau\,e^{\hat{\varphi}_{c}(1)/2}\,\zeta(1)\,\right\}\;.
\end{equation}
Separating the mode relative to the negative eigenvalue $\mu_1$
and proceeding as shown before, we get the following result for
the contribution $Z_0(\eta;A,l\hspace{.04cm})$ of the $m=0$ wave
to
$Z(\eta;A,l\hspace{.04cm})=
e^{-S_{0}(\eta;A,l)/b^2}\prod_{m=0}^{+\infty}
Z_m(\eta;A,l\hspace{.04cm})$
to one loop
\begin{eqnarray}
Z_0(\eta;A,l\hspace{.04cm})&=&\frac{\pi A}{b^2
l^3}\;\frac{1}{(-\,\textrm{det}\,\hat{M}_0)^{1/2}}\;
\frac{\sqrt{2\pi}}{\sqrt{-\mu_1}}\,
\int\hspace{-.06cm}\mathcal{D}\,[\,\zeta_{\perp}\,]\;
e^{-\frac{1}{2}(\zeta_{\perp},D_0\zeta_{\perp})}\nonumber\\
\rule{0pt}{.7cm} &=& \frac{\pi A}{b^2
l^3}\;\frac{1}{(-\,\textrm{det}\,\hat{M}_0)^{1/2}}\;
\frac{1}{(-\textrm{Det}\,D_0)^{1/2}}
\end{eqnarray}
where
\begin{equation}
\textrm{det}\,\hat{M}_0\,=\,
e^{\hat{\varphi}_{c}(1)}\left[\,g_0(1,1) \int_0^1 \int_0^1
e^{\hat{\varphi}_{c}(x)}g_0(x,y)\,e^{\hat{\varphi}_{c}(y)}dx\,dy-
\left(\int_0^1
g_0(x,1)\,e^{\hat{\varphi}_{c}(x)}dx\right)^2\,\right]\;.
\end{equation}
Using the explicit expressions for $e^{\hat{\varphi}_{c}}$ and
$g_0(z,z')$, we get
\begin{equation}
\textrm{det}\,\hat{M}_0\,=\,-\,\frac{(1-a^2)^2}{4(1-2\eta)^4}
\,=\,-\left(\frac{2\pi
A}{l^2}\right)^2\frac{1}{(1-2\eta)^2}\;.
\end{equation}
Summing up, our procedure has lead us to the following expression
\begin{eqnarray}\label{Z(A,l) and f1}
Z(\eta;A,l\hspace{.04cm})&=&e^{-S_{0}(\eta;A,l)/b^2}\;
\frac{(2\pi)^4A}{2\,b^2
l^3}\;
f_1(\eta,a^2)\big(1+O(b^2)\big)\\
\label{unconstrained path int}
\rule{0pt}{.8cm}&=&e^{-S_{0}(\eta;A,l)/b^2}\;\frac{\pi A}{b^2
l^3}\;
\frac{1}{(-\,\textrm{det}\,\hat{M}_0)^{1/2}}\;
\frac{1}{(-\textrm{Det}\,D)^{1/2}}\,\big(1+O(b^2)\big)\\
\label{unconstrained path int bis}
\rule{0pt}{.8cm}&=&e^{-S_{0}(\eta;A,l)/b^2}\;\frac{1-2\eta}{2b^2
l}\; \frac{1}{(-\textrm{Det}\,D)^{1/2}}\,\big(1+O(b^2)\big)
\end{eqnarray}
where the remaining quadratic path integral
$(-\textrm{Det}\,D)^{-1/2}$ involves all the waves $m\geqslant 0$
\begin{equation}
(-\textrm{Det}\,D)^{-1/2}\,=\,\frac{\sqrt{2\pi}}{\sqrt{-\mu_1}}\,
\int\hspace{-.06cm}\mathcal{D}\,[\,\chi_{\perp}\,]\;
e^{-\frac{1}{2}(\chi_{\perp},D\chi_{\perp})}
\end{equation}
and it is unconstrained.

\subsection{The one point function to one loop}
\label{one loop section}

The unconstrained functional integral occurring in
(\ref{unconstrained path int bis}) must be computed with the
boundary conditions
\begin{equation}\label{FZZ b.c. chi disk}
-\,r^2\frac{\partial}{\partial r^2}\,\chi(z)\,=\,\pi
\,\mu_{\scriptscriptstyle\hspace{-.05cm}B}b^2\,
e^{\varphi_{c}/2}\,\chi(z)
\qquad\hspace{.3cm}\textrm{when}\hspace{.3cm}\qquad r^2=1\;.
\end{equation}
To determine the function $f_1(\eta,a^2)\equiv
f(\eta,A/l^2) $ in
(\ref{Z(A,l) and f1}) we shall compute the derivatives of
$\log(-\textrm{Det}\,D)^{-1/2}$ w.r.t. $\eta$ and $a^2$ by
exploiting (\ref{variation logdet FZZ}). Indeed, from
(\ref{g(z,z') equation FZZ}) and (\ref{FZZ b.c. propagator
complex}) one sees that $(-\textrm{Det}\,D)^{-1/2}$ depends only
on $\eta$ and $a^2$. By using the explicit expressions for
$g(z,z)$ and
$g_{\scriptscriptstyle\hspace{-.05cm}B}(\theta,\theta)$ in
(\ref{variation logdet FZZ}), given by (\ref{g(z,z) FZZ}) and
(\ref{g(theta)}) respectively, we find that
\begin{eqnarray}
\left.\frac{\partial}{\partial\eta}\,\log
(-\textrm{Det}\,D)^{-1/2}\,\right|_{a^2}\hspace{-.2cm}&=&
2\,\gamma_{\,\scriptscriptstyle\hspace{-.05cm}E}+\,
\frac{1}{1-2\eta}
\,+\,2\,\psi(1-2\eta)\,-\,2\pi\cot(2\pi\eta) \phantom{xxxxxxx}\\
\rule{0pt}{.9cm} \left.\frac{\partial}{\partial a^2}\,\log
(-\textrm{Det}\,D)^{-1/2}\,\right|_{\eta}\,\,
\hspace{-.2cm}&=&\frac{1}{1-a^2}\;.
\end{eqnarray}
Combining these results, we obtain
\begin{equation}
(-\textrm{Det}\,D)^{-1/2}\,=\, \frac{\beta}{1-a^2}\;
\frac{e^{2\eta\gamma_{\,\scriptscriptstyle\hspace{-.05cm}E}}\,
\Gamma(2\eta)}{\pi\sqrt{1-2\eta}}
\end{equation}
where $\beta$ is a numerical factor.\\
Exploiting the relation (\ref{a^2(A,l)}) and the expression
(\ref{unconstrained path int bis}), the one point function at
fixed area and boundary length reads
\begin{equation}\label{our result}
Z(\eta;A,l\hspace{.04cm})\,=\,e^{-S_{0}(\eta;A,l)/b^2}\;
\frac{\beta}{8\pi^2}
\;\frac{l}{b^2 A}\;
\frac{e^{2\eta\gamma_{\,\scriptscriptstyle\hspace{-.05cm}E}}\,
\Gamma(2\eta)}{\sqrt{1-2\eta}}
\,\big(1+O(b^2)\big)\;.
\end{equation}
The bootstrap approach gives for the one point function at fixed
area and boundary length the following result \cite{FZZ}
\begin{equation}\label{Z bootstrap}
Z_{\eta/b}(A,l\hspace{.04cm})\,=\,\frac{1}{b}\;
\frac{\Gamma(2\eta-b^2)}{\Gamma(1+(1-2\eta)/b^2)}
\left(\frac{l\,\Gamma(b^2)}{2\,A}\right)^{\frac{1-2\eta}{b^2}+1}
\exp\left(-\frac{l^2}{4A\sin(\pi
b^2)}\right)\;.
\end{equation}
The one loop expansion of (\ref{Z bootstrap}) is\footnote{here we
correct a misprint occurring in eq. (2.48) of \cite{FZZ}.}
\begin{eqnarray}
Z_{\eta/b}(A,l\hspace{.04cm})&=&\exp\left\{-\frac{1}{b^2}
\left[\,\frac{l^2}{4\pi
\,A}\,+(1-2\eta)\left(\log\frac{2A}{l}\,+\log(1-2\eta)-1\right)\right]
\right\}\times
\nonumber\\
\rule{0pt}{.9cm}& &
\times\;\frac{
e^{-\gamma_{\,\scriptscriptstyle\hspace{-.05cm}E}}}{2\sqrt{2\pi}}\;
\frac{l}{b^2 A}\;
\frac{e^{2\eta\gamma_{\,\scriptscriptstyle\hspace{-.05cm}E}}\,
\Gamma(2\eta)}{\sqrt{1-2\eta}}
\end{eqnarray}
which agrees with (\ref{our result}), except for the arbitrary
normalization constant $\beta$. Eq. (\ref{our result}) provides
the first perturbative check of the bootstrap result (\ref{Z bootstrap}).\\

\noindent Integrating back (\ref{our result}) in $A$ we obtain
\begin{eqnarray}
\int_{0}^{\infty}\frac{dA}{A}\;e^{-\mu
A}\,Z(\eta;A,l\hspace{.04cm}) &=&
e^{\frac{1-2\eta}{b^2}\,\left(1-\log(1-2\eta)\right)}\,
(\pi\mu b^2)^{\frac{1-2\eta}{2b^2}+\frac{1}{2}}\,\times\\
& & \times\;\frac{\beta}{2\pi^2 b^2}
\;\frac{e^{2\eta\gamma_{\,\scriptscriptstyle\hspace{-.05cm}E}}\,
\Gamma(2\eta)}{\sqrt{1-2\eta}}\;
K_{\frac{1-2\eta}{b^2}+1}\left(\sqrt{\frac{\mu}{\pi
b^2}}\;l\right)
 \,\big(1+O(b^2)\big)\nonumber
\end{eqnarray}
and integrating further this result in $l$ according to
(\ref{one
point fixed A and l def}) we find to one loop
\begin{eqnarray}\label{U one loop}
U(\eta;\mu,\mu_{\scriptscriptstyle\hspace{-.05cm}B})&=&
e^{\frac{1-2\eta}{b^2}\,\left(\frac{1}{2}
\log(\pi\mu b^2)+1-\log(1-2\eta)\right)}\,\times\\
\rule{0pt}{.9cm}& & \hspace{0cm}\times\;\sqrt{\pi\mu
b^2}\;\frac{\beta}{2 \pi b^2} \;
\frac{e^{2\eta\gamma_{\,\scriptscriptstyle\hspace{-.05cm}E}}\,
\Gamma(2\eta)}{\sqrt{1-2\eta}}
\;\frac{\cosh\big(\pi
\sigma\big((1-2\eta)/b^2+1\big)\big)}{\big((1-2\eta)/b^2+1\big)\,
\sin\big(\pi(1-2\eta)/b^2\big)} \nonumber
\end{eqnarray}
where $\sigma$ is defined as follows \cite{FZZ}
\begin{equation}\label{sigma def}
    \big(\cosh(\pi\sigma)\big)^2\,\equiv\,
    \frac{\mu_{\scriptscriptstyle\hspace{-.05cm}B}^2}{\mu}\,\pi
    b^2\;.
\end{equation}
We notice that the factor $1/\sin\big(\pi(1-2\eta)/b^2\big)$,
which displays infinite poles for $b^2\rightarrow 0$, is due to a
divergence at
the origin in the Laplace transform in $l$.\\
The expression (\ref{U one loop}) agrees with the one loop
expansion of the bootstrap formula \cite{FZZ,Teschner: remarks}
\begin{equation}
U(\alpha;\mu,\mu_{\scriptscriptstyle\hspace{-.05cm}B})\,=\,
\frac{2}{b}
\,\big(\pi\mu
\gamma(b^2)\big)^{\frac{Q-2\alpha}{2b}}\,\Gamma(2\alpha
b-b^2)\,\Gamma\left(\frac{2\alpha}{b}-\frac{1}{b^2}-1\right)\,
\cosh\big(\pi s(2\alpha-Q)\big)
\end{equation}
where $Q=1/b+b$, $\gamma(x)=\Gamma(x)/\Gamma(1-x)$ and the
parameter $s$ is defined by
\begin{equation}
\big(\cosh(\pi b
s)\big)^2\,=\,
\frac{\mu_{\scriptscriptstyle\hspace{-.05cm}B}^2}{\mu}\,\sin(\pi
b^2)\;.
\end{equation}
We notice that in the limit $a^2\rightarrow 1$ the semiclassical
contribution to
$U(\eta;\mu,\mu_{\scriptscriptstyle\hspace{-.05cm}B})$ in (\ref{U
one loop}), which is
\begin{equation}ù
e^{\frac{1-2\eta}{b^2}\,\left(\frac{1}{2}\log(\pi\mu
b^2)+1-\log(1-2\eta)-\frac{1}{2}\log
a^2\right)}\,=\,e^{-S_{cl}[\varphi_c]}
\end{equation}
goes over to the semiclassical result of the pseudosphere
\cite{MTpseudosphere}, up to an $\eta$ independent normalization
constant. On the other hand the quantum contribution develops an
infinite number of poles for $b\rightarrow 0$, as discussed after
(\ref{sigma def}).

\section*{Conclusions}

The extension of the technique developed in \cite{MTpseudosphere}
for the pseudosphere has been successfully applied to the
conformal boundary case.\\
A general method has been found for treating functional integrals
with constraints, like the fixed area and boundary length
constraints. We proved that, by properly regularizing the Green
function, the correct quantum dimensions for the vertex functions
are recovered. We gave the explicit computation of the one point
function at fixed area and boundary length to one loop, providing
the first perturbative check of the results obtained through the
bootstrap method \cite{FZZ,Teschner: remarks}.

\section*{Acknowledgments}

We are grateful to Damiano Anselmi, Giovanni Morchio and Massimo
Porrati for useful discussions.

\section*{Appendix}

\appendix
\section{The spectrum of the $D$ operator} \label{spectrum}

\noindent Here we examine the spectrum of the operator
\begin{equation}
\Theta \,\equiv\, \frac{\pi}{2} \,D\,=\,
-\,\partial_z\partial_{\bar z}\,+\, 2\,\pi\mu
b^2\,e^{\varphi_{c}}\,=\, -\,\partial_z\partial_{\bar z}\,+\,
\frac{2\,a^2(1-2\eta)^2}{\big((z\bar{z})^{\eta}
    -a^2(z\bar{z})^{1-\eta}\big)^2}
\end{equation}
with boundary conditions (\ref{FZZ b.c. propagator complex})
\begin{equation}\label{FZZ b.c. appendix z}
\big(\,z\,\partial_z+\bar{z}\,\partial_{\bar{z}}\,\big)\,\chi(z)\,=
\,(1-2\eta)\,\frac{1+a^2}{1-a^2}\;\chi(z)
\qquad\hspace{.3cm}\textrm{when}\hspace{.3cm}\qquad |z|=1
\end{equation}
where $e^{\varphi_{c}}$ is given in (\ref{phiclassic BLFT}).\\
Considering the wave $m=0$, the eigenvalue equation with
eigenvalue $\lambda$
\begin{equation}
\Theta_0\,\chi\,=\,\frac{\pi}{2} \,D_0\,\chi\,=\,\lambda\,\chi
\end{equation}
can be rewritten as
\begin{equation}\label{eigenvalueeq}
-\,(y \,\chi')' + \frac{2}{(1-y)^2}\;\chi\, = y^\rho \Lambda\,\chi
\end{equation}
where $y= a^2(z\bar z)^{1-2\eta}$, $\rho = 2\eta/(1-2\eta)$
and
\begin{equation}
\Lambda = \frac{\lambda}{(1-2\eta)^2 (a^2)^{1/(1-2\eta)}}\;.
\end{equation}
The boundary conditions (\ref{FZZ b.c. appendix z}) read
\begin{equation}\label{FZZbc eigenvalue}
\left.\frac{\chi'}{\chi}\,\right|_{y\,=\,a^2} =\, \frac{1+a^2}{2
a^2(1-a^2)}
\end{equation}
and $\chi(y)$ is regular at the origin. For $\Lambda=0$, the
solution of (\ref{eigenvalueeq}) which is regular at the origin is
\begin{equation}
f_0\,=\,\frac{1+y}{1-y}
\end{equation}
i.e. the function $a_0$ given in (\ref{am functions FZZ}), but it
does not satisfy the boundary conditions (\ref{FZZbc eigenvalue})
because
\begin{equation}
\left.\frac{f'_0}{f_0}\,\right|_{y\,=\,a^2} = \,\frac{2}{1-a^4}
\,<\, \frac{1+a^2}{2 a^2(1-a^2)}
\end{equation}
being $a^2 <1$. Thus we have
\begin{eqnarray}
0\;\;=\;\;\int_{0}^{\,a^2} f_0 (\Theta f_0) \,dy & = &
\int_{0}^{\,a^2} \left(y (f'_0)^2+\frac{2
f_0^2}{(1-y)^2}\right) dy -a^2f'_0(a^2) f_0(a^2)\nonumber\\
\rule{0pt}{.9cm}& = &  \int_{0}^{\,a^2} \left(y (f'_0)^2+\frac{2
f_0^2}{(1-y)^2}\right) dy -\,\frac{2 a^2 (1+a^2)}{(1-a^2)^3}
\end{eqnarray}
i.e.
\begin{equation}
\int_{0}^{\,a^2}\left(y (f'_0)^2+\frac{2 f_0^2}{(1-y)^2}\right) dy
= \frac{2 a^2 (1+a^2)}{(1-a^2)^3}\;.
\end{equation}
Now it is easy to modify slightly $f_0$ near $y=a^2$ to a function
$f_\varepsilon$ satisfying the boundary conditions (\ref{FZZbc
eigenvalue}) and for which
\begin{equation}
\int_{0}^{\,a^2}f_\varepsilon (\Theta f_\varepsilon)\, dy =
\int_{0}^{\,a^2} \left(y (f'_\varepsilon)^2+\frac{2
f^2_\varepsilon}{(1-y)^2}\,\right) dy -a^2 f'_\varepsilon(a^2)
f_\varepsilon(a^2)
\end{equation}
with
\begin{equation}\label{fepsilon boundterm}
\lim_{\varepsilon\rightarrow 0} f'_\varepsilon(a^2)
f_\varepsilon(a^2) = \frac{(1+a^2)^3}{2 a^2(1-a^2)^3}
\end{equation}
and
\begin{equation}\label{fepsilon integral}
\lim_{\varepsilon\rightarrow 0}\,\int_{0}^{\,a^2} \left(y
(f'_\varepsilon)^2+\frac{2 f_\varepsilon^2}{(1-y)^2}\,\right) dy =
\int_{0}^{\,a^2} \left(y (f'_0)+\frac{2 f^2_0}{(1-y)^2}\,\right)
dy =\frac{2 a^2 (1+a^2)}{(1-a^2)^3}\;.
\end{equation}
Being $a^2 <1$, we have that
\begin{equation}
\frac{2 a^2 (1+a^2)}{(1-a^2)^3} <\frac{(1+a^2)^3}{2(1-a^2)^3}
\end{equation}
and therefore on such test function $f_\varepsilon$, which is not
an eigenfunction, we have
\begin{equation}
\int_{0}^{\,a^2} f_\varepsilon \,\Theta f_\varepsilon\, dy \,<\,0
\end{equation}
for sufficiently small $\varepsilon$.
This proves that the operator $\Theta$ is not
positive definite, i.e. it possesses at least one negative
eigenvalue $\lambda_1 <0$.\\
We want now to prove that the ground eigenvalue $\lambda_1$ is the
only negative eigenvalue occurring in the spectrum. First we write
the eigenvalue equation (\ref{eigenvalueeq}) as
\begin{equation}\label{mzeroeq}
(y \chi')'=
\left(\,\frac{2}{(1-y)^2}\,-\,y^\rho\Lambda\right)\chi\;.
\end{equation}
The solution of (\ref{mzeroeq}) which is regular at the origin can
be written as the following convergent series
\begin{eqnarray}\label{serie}
\chi = \chi^{(0)}+\chi^{(1)}+\chi^{(2)}+\dots
\end{eqnarray}
with $\chi^{(0)}=1$ and
\begin{eqnarray}\label{serie nth term}
\chi^{(n)} = \int_0^y(\,\log y - \log
y_1)\left(\,\frac{2}{(1-y_1)^2}\,-\,y_1^\rho\Lambda\right)
\chi^{(n-1)}(y_1)\,dy_1\;.
\end{eqnarray}
From (\ref{serie}) and (\ref{serie nth term}), one immediately
realizes that for $\Lambda<0$ the function $\chi$ is a positive
function, increasing in $y$  and a pointwise increasing function
of $-\Lambda$. Since $\Lambda_1<0$, the ground state eigenfunction
is a positive function. The eigenfunction relative to $\Lambda_2 >
\Lambda_1$ must possess, by orthogonality, at least one node, but,
as we cannot have a node for $\Lambda_2\leqslant 0$, we must have
$\Lambda_2>0$. Thus the operator $\Theta$ with boundary conditions
(\ref{FZZ b.c. appendix z}) has one and only one negative
eigenvalue. The presence of a negative eigenvalue makes
the unconstrained functional integral ill defined.\\
Obviously one has to consider also the positivity of the partial
wave operator for $m=1$ and higher $m$. The eigenvalue equation in
$y=a^2 u = a^2 (z\bar{z})^{1- 2 \eta}$ for $m\geqslant 1$ is
\begin{equation}\label{eigenvalueeq higher m}
-\,(y \chi')' + \,\frac{m^2}{4(1-2
\eta)^2}\;\frac{\chi}{y}\,+\,\frac{2}{(1-y)^2}\;\chi\,=\,
y^\rho\Lambda\,\chi\;.
\end{equation}
It will be sufficient to examine the case $m=1$. The iterative
solution of the following equation
\begin{equation} (y \chi')' -
\,\frac{1}{4(1-2 \eta)^2}\;\frac{\chi}{y}\; =\,
\left(\,\frac{2}{(1-y)^2}\,-y^\rho\Lambda\right)\chi
\end{equation}
is provided by series (\ref{serie}) with
\begin{eqnarray}
\chi^{(0)} &=& y^{\gamma/2} \\
\rule{0pt}{.9cm} \chi^{(n)} &=& \frac{1}{\gamma} \int_0^y
(y^{\gamma/2}y_1^{-\gamma/2}-y^{-\gamma/2}y_1^{\gamma/2})
\left(\,\frac{2}{(1-y_1)^2}\,-y_1^\rho\Lambda\right)
\chi^{(n-1)}(y_1)\,dy_1
\end{eqnarray}
where $\gamma=1/(1-2\eta)$. Since we have always $y_1\leqslant y$,
then
\begin{equation}
y^{\gamma/2}y_1^{-\gamma/2}-y^{-\gamma/2}y_1^{\gamma/2}\,\geqslant\,
0\;.
\end{equation}
Again, being $\chi^{(0)}>0$, we have that the terms of the series
for $\Lambda\leqslant 0$ are positive increasing in $y$  and
pointwise increasing in $-\Lambda$. For $m=1$ and $\eta = 0$ we
know a solution of the equation with null eigenvalue.
It is
\begin{equation}\label{m=1 eta=0 solution}
\chi = \frac{y^{\frac{1}{2}}}{1-y}
\end{equation}
which gives
\begin{equation}
\frac{\chi'}{\chi}=\frac{1+y}{2y(1-y)}
\end{equation}
i.e. it satisfies identically the boundary conditions (\ref{FZZbc
eigenvalue}). Thus for $m=1$ and $\eta=0$ we have the marginal
eigenvalue $\Lambda=0$. Since $\chi$ pointwise increases when
$-\Lambda$ increases, then we cannot have nodes for $\Lambda<0$
and, by orthogonality, we cannot have eigenvalues for $\Lambda<0$
either. Thus, for $m=1$ and $\eta=0$ the operator is positive
semidefinite. Then, from (\ref{eigenvalueeq higher m}), we see
that the operator is positive definite when $m>1$ and
$\eta\geqslant 0$ (always $\eta<1/2$). For $m=1$ and $\eta<0$ the
operator is not positive definite (use as test function the
solution (\ref{m=1 eta=0 solution}) for $m=1$ and $\eta=0$) and
therefore, when $\eta<0$, we have instability also for the $m=1$
wave.

\end{document}